\documentclass[a4paper,12pt]{article}
\usepackage[none]{hyphenat}
\usepackage{setspace}
\usepackage{amsmath}
\usepackage{amsfonts}
\usepackage{mathtools}
\usepackage{amssymb}
\usepackage{tikz}
\usetikzlibrary{shapes.geometric,arrows}
\usepackage{longtable}
\usepackage{float}
\usepackage{pgfplots}
\usepackage{enumitem}
\usepackage{bbm}
\usepackage{multirow}
\usepackage{mathtools}
\usepackage{soul}

\newtheorem{thm}{Theorem}

\newtheorem{cor}{Corollary}
\newtheorem{lem}{Lemma}

\newcommand*{\qed}{\hfill\ensuremath{\square}}

\title{Bonus-malus Systems vs Delays in Claim Settlements: Analysis of Ruin Probabilities}
\author{${}^a$Dhiti Osatakul\thanks{Email: dhiti@cbs.chula.ac.th}, ${}^b$Shuanming Li\thanks{Email: shli@unimelb.edu.au}, ${}^b$Xueyuan Wu\thanks{Corresponding author. Email: xueyuanw@unimelb.edu.au.}\bigskip \\
	{\small ${}^a$Department of Statistics, Chulalongkorn Business School,}\\ 
	{\small Chulalongkorn University, Bangkok, Thailand}\\
	{\small ${}^b$Department of Economics, The University of Melbourne, VIC 3010, AUS} 
	}
\date{}

\begin{document}
\maketitle

\begin{abstract}

Our paper explores a discrete-time risk model with time-varying premiums, investigating two types of correlated claims: main claims and by-claims. Settlement of the by-claims can be delayed for one time period, representing real-world insurance practices. We examine two premium principles based on reported and settled claims, using recursively computable finite-time ruin probabilities to evaluate the performance of time-varying premiums. Our findings suggest that, under specific assumptions, a higher probability of by-claim settlement delays leads to lower ruin probabilities. Moreover, a stronger correlation between main claims and their associated by-claims results in higher ruin probabilities. Lastly, the premium adjustment principles based on settled claims experience contribute to higher ruin probabilities compared to those based on reported claims experience, assuming all other factors remain constant. Notably, this difference becomes more pronounced when there is a high likelihood of by-claim delays.
\vspace{0.5cm}

\end{abstract}

\newpage

\section{Introduction} 

Due to the nature of the insurance business, certain insurers often need to deal with the issue of delayed claim settlements. Many factors prevent insurers from settling claims promptly after the claims are lodged. One of the main causes of delayed claims settlement is the investigation time insurers spend on verifying and assessing the reported claims. A typical example is casualty insurance. According to the usual claiming process of casualty insurance policies, after the policyholders notify the insurance company of the incident that causes loss or damage to their property, the surveyor/loss assessor will detect the reported damage to evaluate the repair/replacement cost. This process may also involve the police department and some third parties, so it may require a lot of time, which results in delayed claims settlement. Another cause of delayed claims settlement is delayed claim reporting. This issue occurs when the policyholder reports a previously incurred insurable loss to the insurer after their insurance policy has expired. In insurance terminology, this type of claim is known as incurred-but-not-reported claims or simply IBNR claims. As the name says, these claims are not reported in a timely manner which certainly delays the whole process of dealing with the claims. In term of the solvency risk, the delayed claims have a significant impact on the loss modelling by actuaries, since the timing of settled claims are inconsistent with the incident occurrence times. It may lead to the underestimation/ overestimation of claim experience in the time period under consideration which will reduce the effectiveness of the insolvency measures developed by usual loss models. Therefore, researchers and practitioners derived methods to deal with delayed or IBNR claims. A well-known approach to dealing with the IBNR claims is the chain ladder method (CLM). It uses the run-off triangles to help insurance companies estimate their required claim reserves involving IBNR losses. In ruin theory, risk models with delayed claims are developed to complement the classical risk model. This type of generalisation relaxes the assumption that claims settlements and claim reporting occur in the same financial period. As a result, the risk models with delayed claims are better connected with real-life insurance practice and attracted much attention from researchers in the literature. \smallskip

Regarding the relevant literature, Waters and Papatriandafylou (1985) derived the upper bounds for the ruin probability of a risk process with delayed claims settlement. Yuen and Guo (2001) studied the ruin probabilities for time-correlated claims in the compound binomial risk model. They introduced the principle of delayed claims in their model by defining the term `main claims', which refers to the initial claims that induce another type of claims, so-called by-claims, with different severity distributions and time occurrence. According to their models, the main claims and by-claims are assumed to be independent, which is a restrictive assumption. Some similar models can be found in Wu and Yuen (2004), which is an extension of Yuen and Guo (2001) by considering the interaction of the dependent classes of business in the models. Xiao and Guo (2007) studied the joint distribution of the surplus immediately prior to ruin and deficit at ruin in the compound binomial risk model with time-correlated claims and its relationship with the classical compound binomial risk model. 

Moreover, Trufin et al. (2011) and  Ahn et al. (2018) studied the ruin probability with IBNR claims. Yuen et al. (2005) applied the martingale theory to obtain the expression for the ultimate ruin probability with the corresponding Lundberg exponent of its non-delayed risk model. Zou and Xie (2010) considered the case that the claims number process follows the Erlang(2) process and derived the explicit expression for the survival probability when both the main claims amount and by-claims amount are exponentially distributed. Dassios and Zhao (2013) obtained an asymptotic expression for the ruin probability with delayed claims by exploiting the non-homogenous Poisson model. Besides, the studies of an approximation of the ruin probability with delayed claims can also be found in Gao et al. (2019) and Yang and Li (2019). For the dividend problem in the risk models with time-delayed claims, Wu and Li (2012) studied the expected present value of dividend payments up to the time of ruin by considering a constant dividend barrier, whereas Zhou et al. (2013) studied a similar problem and assumed that the premium income is governed by the binomial process. Liu and Zhang (2015) considered a randomized dividend strategy for the study of the expected present value of dividend payments up to the time of ruin. Further, the literature concerning the penalty function in the risk models with time-delayed claims can be found in Yuen et al. (2013), Zhu et al. (2014), Liu and Bao (2015), Xie and Zou (2017), Wat et al. (2018), Deng et al. (2018), Zou and Xie (2019) and Liu et al. (2020). 

In this paper, we will extend the study of Yuen and Guo (2001) by assuming that periodic premiums are adjustable and are controlled by previous claims experience. This extension is inspired by the well-known principle in the non-life insurance business, the so-called Bonus-Malus system, which allows the insurers to determine renewal premium levels based on the historical claims experience of the policyholders under consideration. The traditional bonus-malus systems are at the granular level, i.e. at the policyholder level, which ignores the overall financial status of the insurance company. To address the issue, we adopt the portfolio-dependent premium correction framework that is considered crudely, i.e. on the portfolio level or higher, which enables us to incorporate it into the risk models and to study the corresponding ruin probabilities.
As a result, the proposed models in this study can be used to evaluate the risk of ruin for insurers who have to face both delayed claim settlements and varying premiums in their everyday business, such as automobile insurance companies. Studies of risk models with varying premiums can be found in various papers in the literature. For example, Trufin and Loisel (2013) studied the discrete-time risk models with premiums adjusted to the claims by B{\"u}hlmann credibility. Another model in a discrete-time setting can be found in Wu et al. (2015), who used a two-state Markov Chain model to express the ultimate ruin probabilities in terms of both recursive formulae and explicit forms. The related literature regarding the continuous-time setting can be found in Li et al. (2015a) and Constantinescu et al. (2016). In the study of Li et al. (2015a), the premiums were assumed to be adjusted according to the historical claims number, whereas the study of Constantinescu et al. (2016) assumed that the premiums are adjusted according to the change in the inter-arrival time distribution between claims. Additionally, Osatakul and Wu (2021) studied the risk models with claim-dependent premiums and also considered the external environment for their models. Further studies concerning the risk models with varying premiums can also be found in Afonso et al. (2010), Li et al. (2015b), Kucerovský and Najafabadi (2017), Afonso et al. (2017) and Afonso et al. (2020) for the continuous-time setting, and Dufresne (1988) and Wagner (2001) for the discrete-time setting. In this paper, we inherit the assumptions regarding main claims and by-claims from Wu and Li (2012), which weakened the assumptions in Yuen and Guo (2001) by allowing the dependence between main claims and by-claims.\smallskip

It is worth mentioning that if premiums are to be adjusted by the settled claim experience, then the underlying premium status process would display an in-homogeneous nature, because the transition probability between any two premium levels vary from time to time due to the uncertainty in the settled claims. This property differs from the homogeneity property of the premium status process should the premiums be adjusted by the reported claim experience. This interesting contrast makes our discussions in this paper more realistic. 

This paper aims to answer to following questions:
\begin{itemize}
\item What is the impact of the probability of claims settlement delays on the ruin probabilities?
\item What is the impact of the correlation between the main claims and by-claims on the ruin probabilities?
\item Which of the premium adjustment strategies should be implemented by the insurers? In our study, we propose four premium adjustment principles: adjusting by aggregate reported claims, by aggregate settled claims, by reported number of claims and by settled number of claims.  
\end{itemize}\medskip

This paper is organised as follows. Section 2 presents the models and assumptions of our study. Section 3 to 6 presents results for the finite-time ruin probabilities under each of the above four premium adjustment principles respectively. Numerical examples showcasing our theoretical results in this paper are given in Section 7 with detailed discussions.  Concluding remarks and potential future research are given in Section 8.

\section{The Model} 
We first define a surplus process of discrete times, denoted by $U_k$, as
\begin{eqnarray}\label{eq:1}
U_k = U_0 +\sum_{t=1}^{k}(C_t-S_t), \qquad k=0, 1, \ldots,
\end{eqnarray}
where $U_0 \in\,\mathbb{N}:=\{0, 1, 2, \ldots\}$ is the initial surplus, $S_t$ is the total amount of settled claims during the $t$\textsuperscript{th} unit time period payable at time $t$, and $C_t$ is the premium of the $t$\textsuperscript{th} period received at the beginning of the period. In this paper we aim to study varying premiums. Let $\mathbf{c}: = \{c_{1},c_{2},...,c_{l}\}$ be the set of premium levels and $\boldsymbol{\mathcal{L}}=\{1, 2,\ldots,l\}$. Without losing generality, we let $c_1<...<c_l\in\mathbb{N^+}:=\{1, 2, \ldots\}$.

As we mentioned previously, there are two types of reported individual claims, i.e. main claims and the associated by-claims. They are denoted by $X_t$ and $Y_t$ respectively for $t\in\mathbb{N}^+$. In this paper, we only consider a very simple case where there is at most one main claim in any time period, and one main claim generates at most one by-claim. Both $\{X_t\}_{t\in\mathbb{N}^+}$ and $\{Y_t\}_{t\in\mathbb{N}^+}$ are independent and identically distributed (i.i.d.) sequences of random variables with common probability mass function (p.m.f.) $f_X(x), x\in\mathbb{N}$, and $f_Y(y), y\in\mathbb{N}$, respectively. On the other hand, $X_t$ and $Y_t$ are assumed to be correlated with common joint p.m.f. $f_{X Y}(x,y)$, $x,y\in\mathbb{N}$. Not surprisingly, one can see that $f_{X Y}(0,y)=0$ for $y\neq0$. 

Assume that main claims are always settled at the end of the reporting time period, which is not the case for by-claims. When a by-claim $Y_k$ occurs, there is a probability $0\le q\le 1$ that its settlement will be delayed to the end of the $(k+1)$\textsuperscript{th} period. Further, the settlement delays of $Y_k, k=1, 2, \ldots,$ are independent of each other. Thus, the aggregate claim amount settled in time period $t$ is
\begin{equation}
    S_t=
\left\{\begin{array}{l l}
X_t &\mbox{if } Y_t\mbox{ is delayed; no delayed by-claim from } t-1,\\
X_t+Y_{t-1} &\mbox{if } Y_t\mbox{ is delayed; a delayed by-claim from } t-1,\\
X_t+Y_{t} &\mbox{if } Y_t\mbox{ is not delayed; no delayed by-claim from } t-1,\\
X_t+Y_{t}+Y_{t-1} &\mbox{if } Y_t\mbox{ is not delayed; a delayed by-claim from } t-1.
\end{array}\right.\label{eq: S}
\end{equation}

For a given time horizon $n\in\mathbb{N}^+$, the finite-time ruin probability of $U_k$ with initial premium level $c_i$, for $i\in\boldsymbol{\mathcal{L}}$, is defined as
\begin{eqnarray}\label{eq:3}
\psi_i(u,n)=\mathbb{P}_u\Big\{\bigcup\limits_{k=1}^{n}(U_k<0)\Big|C_1=c_i\Big\},
\end{eqnarray}
where the subscript $u$ represents the condition $U_0 = u$. We have $\psi_i(u,n)=1$ for $u<0, n\ge 0$ and $\psi_i(u,0)=0$ for $u\ge 0$ by convention. \smallskip

\textbf{Remark} We assume that there is no delayed by-claim from the time period before the initial time 0. Then, $S_1$ can only be $X_1$ or $X_1+Y_1$. \smallskip

Next we shall develop some recursive algorithm to compute the finite-time ruin probabilities under the above proposed risk framework. To enable our derivations, we define the following auxiliary surplus process with an up-front delayed by-claim \begin{eqnarray}\label{eq:2}
U_k^{'}&=&U_0+\sum_{t=1}^{k}(C_t-S_t)-Y_0,
\end{eqnarray}
where $Y_0>0$ is the up-front delayed by-claim and other notations are exactly the same as those in model \eqref{eq:1}. Assume that $Y_0$ is independent of all other random components in model \eqref{eq:2} and follows the p.m.f. $f_Y(y)$. The corresponding $n$-period finite time ruin probability with initial premium level $c_i$, $i\in\boldsymbol{\mathcal{L}}$, is defined as
\begin{eqnarray}\label{eq:4}
\psi_i^{'}(u;z,n)=\mathbb{P}_u\Big\{\bigcup\limits_{k=1}^{n}(U_k^{'}<0)\Big|C_1=c_i, Y_0=z\Big\}.
\end{eqnarray}
Again, $\psi_i^{'}(u;z,n)=1$ for $u<0, n\ge 0$ and $\psi_i^{'}(u;z,0)=0$ for $u\ge 0$ by convention.\medskip

In the following sections, we shall consider four different premium changing principles, i.e. premiums adjusted according to aggregate reported claims, premiums adjusted according to aggregate settled claims, premiums adjusted according to the reported claim frequency, and premiums adjusted according to the settled claim frequency, respectively.

\section{Premiums adjusted according to aggregate \\reported claims}

The premium changing rule considered in this section allows the next periodic premium to be determined based on the current premium level as well as the total reported claims in the current time period. In our previous model setting, we can see that the total reported claims in time period $k$ is $X_k+Y_k$. Whether the settlement of $Y_k$ is delayed or not does not have impact on the next periodic premium level. We define a bonus-malus system $\Delta = (\mathbf{T},\mathbf{c},i)$, where $i\in\boldsymbol{\mathcal{L}}$ is the state of initial premium level; $\mathbf{T}=\{t_{i j}(s)\}_{i,j\in \boldsymbol{\mathcal{L}}; s\in\mathbb{N}}$ denotes a general set of time-homogeneous rules for premium variations. For any $s\in\mathbb{N}$ and $k\in\mathbb{N}^+$, $t_{ij}(s)=1$ if the total reported claim amount $s$ in time period $k$ leads to the transition from premium level $C_k=c_i$ to $C_{k+1}=c_j$ and $ t_{ij}(s)=0$ otherwise.\smallskip

For any $k\in\mathbb{N}^+$, the probability that the premium level moves from level $c_i$ in time period $k$ to level $c_j$ in time period $k+1$ is defined by 
\begin{eqnarray}\label{eq:2.5}
p_T(i,j)=\sum_{x=0}^{\infty}\sum_{y=0}^{\infty} t_{ij}(x+y) \, f_{X Y}(x,y),\quad \text{for} \quad i,j \in \boldsymbol{\mathcal{L}}.\label{Def:p_T}
 \end{eqnarray}
 
Using \eqref{Def:p_T}, one can obtain a one-step transition probability matrix for the premium level Markov process
\begin{eqnarray*}
{\bold P}_T=[\,p_T(i,j)]\,_{l\times l}=\begin{bmatrix}p_T(1,1)&\cdots&p_T(1,l)\\\vdots&\ddots&\vdots\\p_T(l,1)&\cdots&p_T(l,l)\end{bmatrix}\ .
\end{eqnarray*}

Before we present our first main result, we would like to show a simple relationship between the two finite-time ruin probabilities defined before, which will benefit our following discussions.

\begin{lem} When premiums are adjusted according to aggregate reported claims, the finite-time ruin probabilities $\psi$ and $\psi^{'}$ satisfy the following relationship, for $n\in\mathbb{N}^+$,
\begin{eqnarray}\label{eq:5}
\psi_i^{'}(u;z,n)=\left\{
\begin{array}{l l}
\psi_i(u-z, n) & 0<z\le u,\\
\psi^{'}_i(0; z-u, n) & u<z\le u+c_i,\\
1 & z>u+c_i.
\end{array}\right.
 \end{eqnarray}
\end{lem}

\noindent\textbf{{Proof.}}\\[0.5\baselineskip]
Because the premiums are adjusted according to the total reported claims, the up-front delayed claim $Y_0$ has no impact on how the next premium is going to change. 

When $0<z\leq u$, it can be seen from \eqref{eq:3} that $U'_k$ with initial surplus $u$ is equivalent to $U_k$ with initial surplus $u-z\ge 0$. So the first case of \eqref{eq:5} holds. 

When $z>u+c_i,$ the delayed by-claim is large enough to cause ruin, no matter whether there is any new claim in time period 1.\qed \medskip

Before we present our first main result, we introduce an auxiliary function that is used to simplify our main results given within the rest of this paper: 
$$\xi_y(n):=\sum_{x=1}^nf_{X Y}(x, y+n-x).$$
Our first main result is given below.

\begin{thm} 
Given initial surplus $u\ge 0$ and initial premium level $c_i$, $i\in\boldsymbol{\mathcal{L}}$, the finite-time ruin probability with premiums adjusted according to aggregate reported claims  without the up-front delayed by-claim satisfies the following recursive formula, for $n\in\mathbb{N}^+$,
\begin{eqnarray}\label{eq:7}
\psi_i(u,n+1)&=&\sum_{j=1}^{l}\sum_{x=0}^{u+c_i}\sum_{y=0}^{u+c_i-x}t_{ij}(x+y)\psi_j(u+c_i-x-y,n)f_{X Y}(x,y)\notag\\
&&+\,q\,\sum_{j=1}^{l}\sum_{y=1}^{c_j}t_{ij}(u+c_i+y)\psi^{'}_j(0;y,n)\xi_y(u+c_i)\notag\\
&&+\,q\,\sum_{j=1}^{l}\sum_{y=c_j+1}^{\infty}t_{ij}(u+c_i+y)\xi_y(u+c_i)\notag\\
&&+\,(1-q)\,\sum_{y=1}^{\infty}\xi_y(u+c_i)+\sum_{x=u+c_i+1}^{\infty}f_X(x),
\end{eqnarray}
where $\psi_i(u,1)=(1-q)\,\displaystyle{\sum_{y=1}^{\infty}}\xi_y(u+c_i)+\displaystyle{\sum_{x=u+c_i+1}^{\infty}}f_X(x)$.
\end{thm}

\noindent\textbf{{Proof.}} From \eqref{eq:3}, we have
\begin{eqnarray*}
&&\psi_i(u,n+1)\\
&=&\mathbb{P}_u\Big\{\bigcup\limits_{k=1}^{n+1}(U_k<0)\Big|C_1=c_i\Big\}\\
&=&\sum_{x=u+c_i+1}^{\infty}\mathbb{P}_u\Big\{\bigcup\limits_{k=1}^{n+1}(U_k<0)\Big|C_1=c_i,X_1=x\Big\}f_X(x)\\
&&+\sum_{x=0}^{u+c_i}\sum_{y=0}^{u+c_i-x}\mathbb{P}_u\Big\{\bigcup\limits_{k=2}^{n+1}(U_k<0)\Big|C_1=c_i,X_1=x,Y_1=y\Big\}f_{X Y}(x,y)\\
&&+\,\sum_{x=1}^{u+c_i}\sum_{y=u+c_i-x+1}^{\infty}\bigg[(1-q)\mathbb{P}_u\Big\{U_1<0\Big|C_1=c_i,X_1=x,Y_1=y\Big\}\\
&&\qquad\qquad+q\mathbb{P}_u\Big\{\bigcup\limits_{k=2}^{n+1}(U^{'}_k<0)\Big|C_1=c_i,X_1=x,Y_1=y\Big\}\bigg]f_{X Y}(x,y),
\end{eqnarray*}
where the three major terms within the second equality represent all possibilities of the main claim and by-claim within the first time period. The first term is the scenario that the main claim in the first time period is large enough to cause ruin at time 1. It does not matter whether there is a by-claim or not in this case. The second term covers three scenarios: no claims within the first time period at all, i.e. $X_1=Y_1=0$; only a small by-claim without any by-claims, i.e. $X_1=x, Y_1=0$, $1\le x\le u+c_i$; a small main claim and a small by-claim satisfying $X_1+Y_1\le u+c_i$, where $1\le x\le u+c_i$, $1\le y\le u+c_i-x$. Allowing $x=0$ when $1\le y\le u+c_i-x$ does not hurt as we have made it clear early in Section 2 that $f_{XY}(0,y)=0$ for $y\not= 0$. The third term represents the scenario that there is a small main claim within the first time period paired with a large by-claim satisfying $X_1+Y_1\ge u+c_i+1$, where $1\le x\le u+c_i$, $y\ge u+c_i-x+1$. The two possibilities that the by-claim is settled within this time period or delayed to the next time period are considered separately, where the non-delay case leads to ruin at time 1. Considering the above scenarios and applying the given rule of premium adjustments for the second time period yield
\begin{eqnarray*}
&&\psi_i(u,n+1)\\
&=&\sum_{x=u+c_i+1}^{\infty}f_X(x)+\,(1-q)\,\sum_{x=1}^{u+c_i}\sum_{y=u+c_i-x+1}^{\infty}f_{X Y}(x,y)\\
&&+\sum_{j=1}^{l}\sum_{x=0}^{u+c_i}\sum_{y=0}^{u+c_i-x}t_{ij}(x+y)\mathbb{P}_u\Big\{\bigcup\limits_{k=2}^{n+1}(U_k<0)\Big|C_2=c_j,X_1=x,Y_1=y\Big\}f_{X Y}(x,y)\\
&&+\,q\,\sum_{j=1}^{l}\sum_{x=1}^{u+c_i}\sum_{y=u+c_i-x+1}^{\infty}t_{ij}(x+y)\mathbb{P}_u\Big\{\bigcup\limits_{k=2}^{n+1}(U^{'}_k<0)\Big|C_2=c_j,X_1=x,Y_1=y\Big\}f_{X Y}(x,y)\\
&=&\sum_{x=u+c_i+1}^{\infty}f_X(x)+\,(1-q)\,\sum_{y=1}^{\infty}\xi_y(u+c_i)\\
&&+\sum_{j=1}^{l}\sum_{x=0}^{u+c_i}\sum_{y=0}^{u+c_i-x}t_{ij}(x+y)\mathbb{P}_{u+c_i-x-y}\Big\{\bigcup\limits_{k=2}^{n+1}(U_k<0)\Big|C_2=c_j\Big\}f_{X Y}(x,y)\\
&&+\,q\,\sum_{j=1}^{l}\sum_{x=1}^{u+c_i}\sum_{y=u+c_i-x+1}^{\infty}t_{ij}(x+y)\mathbb{P}_{u+c_i-x}\Big\{\bigcup\limits_{k=2}^{n+1}(U^{'}_k<0)\Big|Y_1=y,C_2=c_j\Big\}f_{X Y}(x,y).
\end{eqnarray*}
Note that we have $U'_k$ in the fourth term of both equalities because there is a delayed by-claim $Y_1$ at the beginning of the second time period. From \eqref{eq:3} and \eqref{eq:4}, we have
\begin{eqnarray*}
&&\psi_i(u,n+1)\\
&=&\sum_{x=u+c_i+1}^{\infty}f_X(x)+(1-q)\,\sum_{y=1}^{\infty}\xi_y(u+c_i)\\
&&+\sum_{j=1}^{l}\sum_{x=0}^{u+c_i}\sum_{y=0}^{u+c_i-x}t_{ij}(x+y)\psi_j(u+c_i-x-y,n)f_{X Y}(x,y)\\
&&+\,q\,\sum_{j=1}^{l}\sum_{x=1}^{u+c_i}\sum_{y=u+c_i-x+1}^{u+c_i+c_j-x}t_{ij}(x+y)\psi^{'}_j(u+c_i-x;y,n)f_{X Y}(x,y)\\
&&+\,q\,\sum_{j=1}^{l}\sum_{x=1}^{u+c_i}\sum_{y=u+c_i+c_j-x+1}^{\infty}t_{ij}(x+y)f_{X Y}(x,y)\\
&=&\sum_{j=1}^{l}\sum_{x=0}^{u+c_i}\sum_{y=0}^{u+c_i-x}t_{ij}(x+y)\psi_j(u+c_i-x-y,n)f_{X Y}(x,y)\\
&&+\,q\,\sum_{j=1}^{l}\sum_{y=1}^{c_j}t_{ij}(u+c_i+y)\psi^{'}_j(0;y,n)\xi_y(u+c_i+y)+\sum_{x=u+c_i+1}^{\infty}f_X(x)\\
&&+\,q\,\sum_{j=1}^{l}\sum_{y=c_j+1}^{\infty}t_{ij}(u+c_i+y)\xi_y(u+c_i+y)+\,(1-q)\,\sum_{y=1}^{\infty}\xi_y(u+c_i).
\end{eqnarray*}
One can also verify that 
\begin{eqnarray*}
\psi_i(u,1)&=&\mathbb{P}_u\Big\{U_1<0\Big|C_1=c_i\Big\}\\
&=&\sum_{x=u+c_i+1}^{\infty}\mathbb{P}_u\Big\{U_1<0\Big|C_1=c_i,X_1=x\Big\}f_X(x)\\
&&+\sum_{x=0}^{u+c_i}\sum_{y=0}^{u+c_i-x}\mathbb{P}_u\Big\{U_1<0\Big|C_1=c_i,X_1=x,Y_1=y\Big\}f_{X Y}(x,y)\\
&&+\,\sum_{x=1}^{u+c_i}\sum_{y=u+c_i-x+1}^{\infty}\bigg[(1-q)\mathbb{P}_u\Big\{U_1<0\Big|C_1=c_i,X_1=x,Y_1=y\Big\}\\
&&\qquad\qquad+q\mathbb{P}_u\Big\{U_1<0\Big|C_1=c_i,X_1=x,Y_1=y\Big\}\bigg]f_{X Y}(x,y)\\
&=&\,(1-q)\,\sum_{y=1}^{\infty}\xi_y(u+c_i)+\sum_{x=u+c_i+1}^{\infty}f_X(x).
\end{eqnarray*}\qed \\

\noindent {\bf Remark.} From the definition of $\xi_y(n+y)$, one can show that 
\begin{eqnarray*}
\sum^\infty_{y=1}\xi_y(n)&=&\sum^n_{x=1}\sum^\infty_{y=1}f_{X Y}(x, n-x+y)\\
&=&\sum^n_{x=1}\Big[\sum^\infty_{y=0}f_{X Y}(x, y)-\sum^{n-x}_{y=0}f_{X Y}(x, y)\Big]\\
&=&\sum^n_{x=1}\Big[f_X(x)-\sum^{n-x}_{y=0}f_{X Y}(x, y)\Big].
\end{eqnarray*}
Also, $\sum_{x=u+c_i+1}^{\infty}f_X(x)=1-\sum_{x=0}^{u+c_i}f_X(x)$. Therefore, in the recursive formula given in Theorem 1, there is only one infinite summation left which requires extra attention when use it for computational purposes.\medskip

To use the recursive formula obtained in Theorem 1, we need to find a way to determine $\psi^{'}_i(0; z, n)$, $0<z\le u+c_i$, $n\in\mathbb{N}^+$.

\begin{cor}
The finite-time ruin probability with premiums adjusted according to aggregate reported claims and an up-front delayed by-claim $z$ satisfies the following recursive formula, for $0<z\leq u+c_i$ and $n\in\mathbb{N}^+$,
\begin{eqnarray}\label{eq:8}
\psi^{'}_i(0;z,n+1)&=&\sum_{j=1}^{l}\sum_{x=0}^{c_i-z}\sum_{y=0}^{c_i-z-x}t_{ij}(x+y)\psi_j(c_i-z-x-y,n)f_{X Y}(x,y)\notag\\
&&+\,q\,\sum_{j=1}^{l}\sum_{y=1}^{c_j}t_{ij}(c_i-z+y)\psi^{'}_j(0;y,n)\xi_y(c_i-z)\notag\\
&&\,q\,\sum_{j=1}^{l}\sum_{y=c_j+1}^{\infty}t_{ij}(c_i-z+y)\xi_y(c_i-z)\notag\\
&&+(1-q)\,\sum_{y=1}^{\infty}\xi_y(c_i-z)+\sum_{x=c_i-z+1}^{\infty}f_X(x),
\end{eqnarray}
where $\psi^{'}_i(0;z,1)=(1-q)\,\sum\limits_{y=1}^{\infty}\xi_y(c_i-z)+\displaystyle{\sum_{x=c_i-z+1}^{\infty}}f_X(x)$.
\end{cor}

\noindent\textbf{{Proof.}}  For $u<z\le u+c_i$, the same method in the proof of Theorem 1 can be used to derive a recursive formula for $\psi^{'}_i(u;z,n+1)$. Using \eqref{eq:5}, \eqref{eq:8} can be obtained by replacing $\psi^{'}_i(u;z,n)$ with $\psi^{'}_i(0;z-u,n)$ in the formula.\qed 

\section{Premiums adjusted according to aggregate \\settled claims}

Previously, we have discussed the first case of varying premiums based on the total reported claims. In contrast, we shall consider another case where for $k\in\mathbb{N}^+$, the premium $C_{k+1}$ is determined by $C_k$ and the total settled claims in time period $k$, i.e. $S_k$. Other model assumptions are the same as the previous case. \medskip

It is worth noting that in this case of premium correction, the underlying Markov process governing the periodic premium levels is not time-homogeneous anymore since the distribution of aggregate settled claims $S_t$ takes different forms over time, see \eqref{eq: S} for details. Further, Lemma 1 does not hold in this case either as having a by-claim delayed from previous time period or not does matter when determining future premiums. However, we can still follow the main idea in previous section to obtain the following main result. 

\begin{thm}
Given initial surplus $u\ge 0$ and initial premium level $c_i$, $i\in\boldsymbol{\mathcal{L}}$, the finite-time ruin probability with premiums adjusted according to aggregate settled claims without the up-front delayed by-claim satisfies the following recursive formula, for $n\in\mathbb{N}^+$,
\begin{eqnarray}\label{eq:9}
\psi_i(u,n+1)&=&\sum_{j=1}^{l}\sum_{x=0}^{u+c_i}t_{ij}(x)\psi_j(u+c_i-x,n)f_{X Y}(x,0)+(1-q)\sum_{y=1}^{\infty}\xi_y(u+c_i)\notag\\
&&+\,(1-q)\,\sum_{j=1}^{l}\sum_{x=1}^{u+c_i-1}\sum_{y=1}^{u+c_i-x}t_{ij}(x+y)\psi_j(u+c_i-x-y,n)f_{X Y}(x,y)\notag\\
&&+\,q\,\sum_{j=1}^{l}\sum_{x=1}^{u+c_i}\sum_{y=1}^{u+c_i-x+c_j}t_{ij}(x)\psi^{'}_j(u+c_i-x;y,n)f_{X Y}(x,y)\notag\\
&&+\,q\,\sum_{j=1}^{l}\sum_{x=1}^{u+c_i}\sum_{y=u+c_i-x+c_j+1}^{\infty}t_{ij}(x)f_{X Y}(x,y)
+\sum_{x=u+c_i+1}^{\infty}f_X(x),
\end{eqnarray}
where $\psi_i(u,1)=\displaystyle{\sum_{x=u+c_i+1}^{\infty}}f_X(x)+(1-q)\,\sum_{y=1}^{\infty}\xi_y(u+c_i)$.
\end{thm}

\noindent\textbf{{Proof.}} From \eqref{eq:3}, we have
\begin{eqnarray*}
&&\psi_i(u,n+1)\\
&=&\mathbb{P}_u\Big\{\bigcup\limits_{k=1}^{n+1}(U_k<0)\Big|C_1=c_i\Big\}\\
&=&\sum_{x=0}^{u+c_i}\sum_{y=0}^{\infty}\mathbb{P}_u\Big\{\bigcup\limits_{k=1}^{n+1}(U_k<0)\Big|C_1=c_i,X_1=x,Y_1=y\Big\}f_{X Y}(x,y)\\
&&+\sum_{x=u+c_i+1}^{\infty}\mathbb{P}_u\Big\{\bigcup\limits_{k=1}^{n+1}(U_k<0)\Big|C_1=c_i,X_1=x\Big\}f_X(x)\\
&=&\sum_{x=0}^{u+c_i}\mathbb{P}_u\Big\{\bigcup\limits_{k=1}^{n+1}(U_k<0)\Big|C_1=c_i,X_1=x,Y_1=0\Big\}f_{X Y}(x,0)\\
&&+\sum_{x=1}^{u+c_i}\sum_{y=1}^{\infty}\bigg[(1-q)\,\mathbb{P}_u\Big\{\bigcup\limits_{k=1}^{n+1}(U_k<0)\Big|C_1=c_i,X_1=x,Y_1=y\Big\}\\
&&+q\,\mathbb{P}_u\Big\{\bigcup\limits_{k=2}^{n+1}(U^{'}_k<0)\Big|C_1=c_i,X_1=x,Y_1=y\Big\}\bigg]f_{X Y}(x,y)\\
&&+\sum_{x=u+c_i+1}^{\infty}f_X(x).
\end{eqnarray*}
The scenarios listed in the third equality are the same as those in the second equality within the proof of Theorem 1. Then we have
\begin{eqnarray*}
&&\psi_i(u,n+1)\\
&=&\sum_{x=0}^{u+c_i}\mathbb{P}_u\Big\{\bigcup\limits_{k=1}^{n+1}(U_k<0)\Big|C_1=c_i,X_1=x,Y_1=0\Big\}f_{X Y}(x,0)\\
&&+\,(1-q)\,\bigg[\sum_{x=1}^{u+c_i-1}\sum_{y=1}^{u+c_i-x}\mathbb{P}_u\Big\{\bigcup\limits_{k=1}^{n+1}(U_k<0)\Big|C_1=c_i,X_1=x,Y_1=y\Big\}f_{X Y}(x,y)\\
&&+\sum_{x=1}^{u+c_i}\sum_{y=u+c_i-x+1}^{\infty}f_{X Y}(x,y)\bigg]+\sum_{x=u+c_i+1}^{\infty}f_X(x)\\
&&+\,q\,\sum_{x=1}^{u+c_i}\sum_{y=1}^{\infty}\mathbb{P}_u\Big\{\bigcup\limits_{k=2}^{n+1}(U^{'}_k<0)\Big|C_1=c_i,X_1=x,Y_1=y\Big\}f_{X Y}(x,y)
\end{eqnarray*}

\begin{eqnarray*}
&=&\sum_{j=1}^{l}\sum_{x=0}^{u+c_i}t_{ij}(x)\mathbb{P}_{u+c_i-x}\Big\{\bigcup\limits_{k=2}^{n+1}(U_k<0)\Big|C_2=c_j\Big\}f_{X Y}(x,0)\\
&&+\,(1-q)\,\sum_{j=1}^{l}\sum_{x=1}^{u+c_i-1}\sum_{y=1}^{u+c_i-x}t_{ij}(x+y)\mathbb{P}_{u+c_i-x-y}\Big\{\bigcup\limits_{k=2}^{n+1}(U_k<0)\Big|C_2=c_j\Big\}f_{X Y}(x,y)\\
&&+(1-q)\sum_{x=1}^{u+c_i}\sum_{y=u+c_i-x+1}^{\infty}f_{X Y}(x,y)+\sum_{x=u+c_i+1}^{\infty}f_X(x)\\
&&+\,q\,\sum_{j=1}^{l}\sum_{x=1}^{u+c_i}\sum_{y=1}^{\infty}t_{ij}(x)\mathbb{P}_{u+c_i-x}\Big\{\bigcup\limits_{k=2}^{n+1}(U^{'}_k<0)\Big|Y_1=y,C_2=c_j\Big\}f_{X Y}(x,y)\\
&=&\sum_{j=1}^{l}\sum_{x=0}^{u+c_i}t_{ij}(x)\psi_j(u+c_i-x,n)f_{X Y}(x,0)+(1-q)\sum_{y=1}^{\infty}\xi_y(u+c_i)\\
&&+\,(1-q)\,\sum_{j=1}^{l}\sum_{x=1}^{u+c_i-1}\sum_{y=1}^{u+c_i-x}t_{ij}(x+y)\psi_j(u+c_i-x-y,n)f_{X Y}(x,y)\\
&&+\,q\,\sum_{j=1}^{l}\sum_{x=1}^{u+c_i}\sum_{y=1}^{u+c_i-x+c_j}t_{ij}(x)\psi^{'}_j(u+c_i-x;y,n)f_{X Y}(x,y)\\
&&+\,q\,\sum_{j=1}^{l}\sum_{x=1}^{u+c_i}\sum_{y=u+c_i-x+c_j+1}^{\infty}t_{ij}(x)f_{X Y}(x,y)
+\sum_{x=u+c_i+1}^{\infty}f_X(x).
\end{eqnarray*}
Note that the rule of premium adjustments applied above is different from the rule proposed in Section 3. Here only the settled claims are counted when analysing the aggregate claims for the premium adjustment purposes. Similar to Theorem 1, one can verify that the result for $\psi_i(u, 1)$ is just a special case of $n=1$.\qed \\

To use the recursive formula obtained in Theorem 2, we need to find a way to determine $\psi^{'}_i(u; z, n)$, $0<z\le c_i$, $n\in\mathbb{N}^+$.

\begin{cor}
The finite-time ruin probability with premiums adjusted according to aggregate settled claims and an up-front delayed by-claim $z$ satisfies the following recursive formula, for $0<z\le u+c_i$ and $n\in\mathbb{N}^+$,
\begin{eqnarray}\label{eq:10}
\psi^{'}_i(u;z,n+1)&=&\,\sum_{j=1}^{l}\sum_{x=0}^{u-z+c_i}t_{ij}(x+z)\psi_j(u-z+c_i-x,n)f_{X Y}(x,0)\notag\\
&&+\,(1-q)\,\sum_{j=1}^{l}\sum_{x=1}^{u-z+c_i-1}\sum_{y=1}^{u-z+c_i-x}t_{ij}(x+y+z)\nonumber\\
&&\times\psi_j(u-z+c_i-x-y,n)f_{X Y}(x,y)\nonumber\\
&&+\,q\,\sum_{j=1}^{l}\sum_{x=1}^{u-z+c_i}\sum_{y=1}^{u-z+c_i-x+c_j}t_{ij}(x+z)\psi^{'}_j(u-z+c_i-x;y,n)f_{X Y}(x,y)\nonumber\\
&&+\,q\,\sum_{j=1}^{l}\sum_{x=1}^{u-z+c_i}\sum_{y=u-z+c_i-x+c_j+1}^{\infty}t_{ij}(x+z)f_{X Y}(x,y)\nonumber\\
&&+(1-q)\sum_{y=1}^{\infty}\xi_y(u-z+c_i)+\sum_{x=u-z+c_i+1}^{\infty}f_X(x),
\end{eqnarray}
where $\psi^{'}_i(u;z,1)=\displaystyle{\sum_{x=u-z+c_i+1}^{\infty}}f_X(x)+(1-q)\,\sum_{y=1}^{\infty}\xi_y(u-z+c_i)$.
\end{cor}

\noindent\textbf{{Proof.}} Using \eqref{eq:9}, \eqref{eq:10} can be obtained by adding $z$ into the premium rule function $t_{ij}$  and replacing $u$ by $u-z$ in \eqref{eq:9}.\qed \\

\section{Premiums adjusted according to reported number of claims} 

In this section, we shall switch the premium correction trigger from aggregate claim experience to claim frequency experience. We still denote the bonus-malus system by $\Delta = (\mathbf{T},\mathbf{c},i)$, where $i\in\boldsymbol{\mathcal{L}}$; $\mathbf{T}=\{t_{i j}(k)\}_{i,j\in \boldsymbol{\mathcal{L}}; k\in\mathbb{N}}$ denotes a general set of time-homogeneous rules with input $k$ being the number of claims. For any $k\in\mathbb{N}$ and $n\in\mathbb{N}^+$, $t_{ij}(k)=1$ if the total number of claims in time period $n$ leads to the transition from premium level $C_n=c_i$ to $C_{n+1}=c_j$ and $ t_{ij}(k)=0$ otherwise.

Now we consider the first type of claim frequency, i.e. the total number of reported claims. Let $N^X_t$ denote the number of main claims in $t$-th time period. According to the assumption in section 2, we have $\mathbb{P}(N^X_t=0)=f_X(0)$ and $\mathbb{P}(N^X_t=1)=1-f_X(0)$. Similarly, the number of by-claims in time period $t$ is denoted by $N^Y_t$, where $\mathbb{P}(N^Y_t=0)=f_Y(0)$ and $\mathbb{P}(N^Y_t=1)=1-f_Y(0)$. We assume that $N^Y_t$ is observable at time $t$ no matter if the settlement  of $Y_t$ will be delayed or not, so the total number of reported claims in time period $t$ is $N^X_t+N^Y_t$. For any time period $t$, $t\in\mathbb{N}^+$, there are only three cases of reported number of claims:\\
\begin{enumerate}[label=\arabic*),nolistsep]
\item $N^X_t=0$ and $N^Y_t=0$;
\item $N^X_t=1$ and $N^Y_t=0$;
\item $N^X_t=1$ and $N^Y_t=1$.
\end{enumerate}

\medskip
Using a similar method as the one used in Section 3, we obtain the following main result:

\begin{thm} 
Given initial surplus $u\ge 0$ and initial premium level $c_i$, $i\in\boldsymbol{\mathcal{L}}$, the finite-time ruin probability with premiums adjusted according to reported number of claims without the up-front delayed by-claim satisfies the following recursive formula, for $n\in\mathbb{N}^+$,
\begin{eqnarray}\label{eq:11}
\psi_i(u,n+1)&=&\sum_{j=1}^{l}t_{ij}(0)\psi_j(u+c_i,n)f_{X Y}(0,0)\notag\\
&&+\sum_{j=1}^{l}t_{ij}(1)\sum_{x=1}^{u+c_i}\psi_j(u+c_i-x,n)f_{X Y}(x,0)\notag\\
&&+\sum_{j=1}^{l}t_{ij}(2)\bigg(\sum_{x=1}^{u+c_i-1}\sum_{y=1}^{u+c_i-x}\psi_j(u+c_i-x-y,n)f_{X Y}(x,y)\notag\\
&&+\,q\,\sum_{y=1}^{c_j}\psi^{'}_j(0;y,n)\xi_y(u+c_i)+\,q\,\sum_{y=c_j+1}^{\infty}\xi_y(u+c_i)\bigg)\notag\\
&&+\,(1-q)\,\sum_{y=1}^{\infty}\xi_y(u+c_i)+\sum_{x=u+c_i+1}^{\infty}f_X(x),
\end{eqnarray}
where $\psi_i(u,1)=(1-q)\,\displaystyle{\sum_{y=1}^{\infty}}\xi_y(u+c_i)+\displaystyle{\sum_{x=u+c_i+1}^{\infty}}f_X(x)$.
\end{thm}

\noindent {\bf Proof.} It is not hard to see that the formula \eqref{eq:11} can be obtained by replacing the premium rule $t_{ij}$ in \eqref{eq:7} with the new version defined at the beginning of this section. Since there are only three cases of total number of reported claims in each time period, one can get \eqref{eq:11} straightforwardly. \hfill $\Box$

 \begin{cor}
The finite-time ruin probability with premiums adjusted according to reported claims number and an up-front delayed by-claim $z$ satisfies the following recursive formula, for $0<z\leq c_i$ and $n\in\mathbb{N}^+$,
\begin{eqnarray}\label{eq:12}
\psi^{'}_i(0;z,n+1)&=&\sum_{j=1}^{l}t_{ij}(0)\psi_j(c_i-z,n)f_{X Y}(0,0)\notag\\
&&+\sum_{j=1}^{l}t_{ij}(1)\sum_{x=1}^{c_i-z}\psi_j(c_i-z-x,n)f_{X Y}(x,0)\notag\\
&&+\sum_{j=1}^{l}t_{ij}(2)\bigg(\sum_{x=1}^{c_i-z-1}\sum_{y=1}^{c_i-z-x}\psi_j(c_i-z-x-y,n)f_{X Y}(x,y)\notag\\
&&+\,q\,\sum_{y=1}^{c_j}\psi^{'}_j(0;y,n)\xi_y(c_i-z)+\,q\,\sum_{y=c_j+1}^{\infty}\xi_y(c_i-z)\bigg)\notag\\
&&+\,(1-q)\,\sum_{y=1}^{\infty}\xi_y(c_i-z)+\sum_{x=c_i-z+1}^{\infty}f_X(x),
\end{eqnarray}
where $\psi^{'}_i(0;z,1)=(1-q)\,\displaystyle{\sum_{y=1}^{\infty}\xi_y(c_i-z)}+\displaystyle{\sum_{x=c_i-z+1}^{\infty}}f_X(x)$.
\end{cor}

\noindent {\bf Proof.} Again, the formula \eqref{eq:12} can be obtained by plugging in the reported claims number in the new premium rule function $t_{ij}$ that replaces the one in \eqref{eq:8}. \hfill $\Box$

\section{Premiums adjusted according to settled number of claims}

In this section, we consider the second type of claim frequency, i.e. the total number of settled claims in a given time period. For time period $t\in\mathbb{N}^+$, let $M^X_t$ denote the number of main claims in this time period; let $M^Y_t$ denote the number of settled by-claims incurred in the current period; let $M^Z_t$ denote the number of settled by-claims incurred in precious time period. Based on the assumptions in Section 2, we know that all of these count random variables can only take a value either 0 or 1. The values of $M^Z_t$ and $M^X_t$ have unique interpretations, but the value 0 for $M^Y_t$ leads to multiple possibilities. To be more specific, $M^Y_t=0$ means either no by-claim incurred in time period $t$ or the settlement of the incurred by-claim is delayed to next time period. This implies that $M^Z_{t+1}=1$ gives $M^Y_t=0$, but not vice versa.

Here we assume that the premium $C_{t+1}$, $t\in\mathbb{N}^+$, is determined according to the total number of settled claims in time period $t$, i.e. $M^X_t+M^Y_t+M^Z_t$, which can take an integer value from 0 to 3:\medskip
\begin{enumerate}[label=\arabic*),nolistsep]
\item $M^Z_t=0$, $M^X_t=0$, $M^Y_t=0$ $\Rightarrow$ $M^X_t+M^Y_t+M^Z_t=0$;
\item $M^Z_t=0$, $M^X_t=1$, $M^Y_t=0$ $\Rightarrow$ $M^X_t+M^Y_t+M^Z_t=1$;
\item $M^Z_t=0$, $M^X_t=1$, $M^Y_t=1$ $\Rightarrow$ $M^X_t+M^Y_t+M^Z_t=2$;
\item $M^Z_t=1$, $M^X_t=0$, $M^Y_t=0$ $\Rightarrow$ $M^X_t+M^Y_t+M^Z_t=1$;
\item $M^Z_t=1$, $M^X_t=1$, $M^Y_t=0$ $\Rightarrow$ $M^X_t+M^Y_t+M^Z_t=2$;
\item $M^Z_t=1$, $M^X_t=1$, $M^Y_t=1$ $\Rightarrow$ $M^X_t+M^Y_t+M^Z_t=3$.
\end{enumerate}\medskip

Similar to Section 4, there is lack of time-homogeneity in the underlying Markov process for premiums. Taking into account the complications illustrated above on the total number of settled claims, we obtain the following result for the finite-time ruin probabilities.

\begin{thm}
Given initial surplus $u\ge 0$ and initial premium level $c_i$, $i\in\boldsymbol{\mathcal{L}}$, the finite-time ruin probability with premiums adjusted according to settled claims number without the up-front delayed by-claim satisfies the following recursive formula, for $n\in\mathbb{N}^+$,
\begin{eqnarray}\label{eq:13}
\psi_i(u,n+1)&=&\,\sum_{j=1}^{l}t_{ij}(0)\psi_j(u+c_i,n)f_{X Y}(0,0)\notag\\
&&+\,\sum_{j=1}^{l}\sum_{x=1}^{u+c_i}t_{ij}(1)\psi_j(u+c_i-x,n)f_{X Y}(x,0)\notag\\
&&+\,(1-q)\,\sum_{j=1}^{l}\sum_{x=1}^{u+c_i-1}\sum_{y=1}^{u+c_i-x}t_{ij}(2)\psi_j(u+c_i-x-y,n)f_{X Y}(x,y)\nonumber\\
&&+\,q\,\sum_{j=1}^{l}\sum_{x=1}^{u+c_i}\sum_{y=1}^{u+c_i-x+c_j}t_{ij}(1)\psi^{'}_j(u+c_i-x;y,n)f_{X Y}(x,y)\nonumber\\
&&+\,q\,\sum_{j=1}^{l}\sum_{x=1}^{u+c_i}\sum_{y=u+c_i-x+c_j+1}^{\infty}t_{ij}(1)f_{X Y}(x,y)\nonumber\\
&&+(1-q)\sum_{y=1}^{\infty}\xi_y(u+c_i)+\sum_{x=u+c_i+1}^{\infty}f_X(x),
\end{eqnarray}
where $\psi_i(u,1)=\displaystyle{\sum_{x=u+c_i+1}^{\infty}}f_X(x)+(1-q)\,\sum_{y=1}^{\infty}\xi_y(u+c_i)$.
\end{thm}

\noindent {\bf Proof.} Similar to the proof of Theorem 3, \eqref{eq:13} can be obtained by replacing the function $t_{ij}$ in \eqref{eq:9} with the new one defined at the beginning of Section 5. Then we plug in the total number of settled claims in $t_{ij}$ according to the values of $X$ and $Y$. \hfill $\Box$

\begin{cor}
The finite-time ruin probability with premiums adjusted according to settled claims number and an up-front delayed by-claim $z$ satisfies the following recursive formula, for $0<z\le u+c_i$ and $n\in\mathbb{N}^+$,

\begin{eqnarray}\label{eq:14}
\psi^{'}_i(u;z,n+1)&=&\,\sum_{j=1}^{l}t_{ij}(1)\psi_j(u-z+c_i,n)f_{X Y}(0,0)\notag\\
&&+\,\sum_{j=1}^{l}\sum_{x=1}^{u-z+c_i}t_{ij}(2)\psi_j(u-z+c_i-x,n)f_{X Y}(x,0)\notag\\
&&+\,(1-q)\,\sum_{j=1}^{l}\sum_{x=1}^{u-z+c_i-1}\sum_{y=1}^{u-z+c_i-x}t_{ij}(3)\psi_j(u-z+c_i-x-y,n)f_{X Y}(x,y)\nonumber\\
&&+\,q\,\sum_{j=1}^{l}\sum_{x=1}^{u-z+c_i}\sum_{y=1}^{u-z+c_i-x+c_j}t_{ij}(2)\psi^{'}_j(u-z+c_i-x;y,n)f_{X Y}(x,y)\nonumber\\
&&+\,q\,\sum_{j=1}^{l}\sum_{x=1}^{u-z+c_i}\sum_{y=u-z+c_i-x+c_j+1}^{\infty}t_{ij}(2)f_{X Y}(x,y)\nonumber\\
&&+(1-q)\sum_{y=1}^{\infty}\xi_y(u-z+c_i)+\sum_{x=u-z+c_i+1}^{\infty}f_X(x),
\end{eqnarray}
where $\psi^{'}_i(u;z,1)=\displaystyle{\sum_{x=u-z+c_i+1}^{\infty}}f_X(x)+(1-q)\,\sum_{y=1}^{\infty}\xi_y(u-z+c_i)$.
\end{cor}
\noindent\textbf{{Proof.}} Using \eqref{eq:13}, \eqref{eq:14} can be obtained by adding $1$ into the premium rule function $t_{ij}$  and replacing $u$ by $u-z$ in \eqref{eq:13}.\qed

\section{Numerical results}

In this section we shall provide some numerical examples to illustrate the theoretical results obtained under the previously discussed four premium adjustment principles and to further study the commonality and dissimilarity of the four principles. Since we have been focusing on the finite-time ruin probabilities in this paper, we shall adopt the finite-time ruin probabilities with a fixed term (say 20) as the proxy to achieve the aforementioned goals. The possible behaviours of finite-time ruin probabilities under each principle when the term changes are not covered here, mainly due to the significantly increased computational costs involved in the completion of the task. 

\subsection{Premiums adjusted according to aggregate reported claims}
The first numerical example we give in this section applies the premium correction principle allowing premiums to be adjusted according to aggregate reported claims. As mentioned previously in section 2.1, the aggregate claims are assumed to be reported at the end of each policy period even when the settlement of by-claims is delayed. We shall examine three hypothetical scenarios for the degree of correlation between the main claim $X$ and the by-claim $Y$ in this example: low correlation, moderate correlation and high correlation. For each scenario, two cases of claim settlement delay are considered: $q=0.2$ or $q=0.8$. We propose the following joint distributions of $X$ and $Y$: 
\begin{itemize}
    \item high correlation case:
\[
    f^{H}_{X Y}(x,y)= 
\begin{cases}
\frac{1}{6}& x=y=0,\\
 \Big(\frac{1}{6}\Big)\Big(\frac{5}{6}\Big)^x& x=y>0,\\
 0 & \text{otherwise},
\end{cases}
\]
where $\mathbb{E}(X)$=5, $\mathbb{E}(Y)$=5 and the correlation coefficient $\rho_{X Y}$=1;
\item low correlation case: 
\[
    f^{L}_{X Y}(x,y)= 
\begin{cases}
\frac{1}{6}& x=y=0,\\
 \Big(\frac{1}{6}\Big)\Big(\frac{5}{6}\Big)^x \Big(\frac{1}{7}\Big)\Big(\frac{6}{7}\Big)^y& x>0,\, y\geq0 ,\\
0 & \text{otherwise},
\end{cases}
\]
where $\mathbb{E}(X)$=5, $\mathbb{E}(Y)$=5 and  $\rho_{X Y}$=0.1443;
\item moderate correlation case: we let $$ f^{M}_{X Y}(x,y)=0.5\;f^{H}_{X Y}(x,y)+0.5\;f^{L}_{X Y}(x,y),$$ where  $\mathbb{E}(X)$=5, $\mathbb{E}(Y)$=5 and so $\rho_{X Y}$=0.5401.
\end{itemize}
According to the above assumptions, we can see that $X$ follows the same marginal geometric distribution in all three cases, i.e. $f_X(x)=\Big(\frac{1}{6}\Big)\Big(\frac{5}{6}\Big)^x$, $x\ge 0$.
However, the three marginal distributions of $Y$ differ from each other, which are listed below, for $y\ge 0$,
\begin{eqnarray*}
&&f_Y^H(y)= \big(\tfrac{1}{6}\big)\big(\tfrac{5}{6}\big)^y; \\
&&f_Y^L(y)=\tfrac{1}{6}\times I_{\{y=0\}}+\tfrac{5}{6}\times\big(\tfrac{1}{7}\big)\big(\tfrac{6}{7}\big)^y;\\
&&f_Y^M(y)=\tfrac{1}{2}\,f_Y^H(y)+\tfrac{1}{2}\,f_Y^L(y),
\end{eqnarray*}
where $I_{\{y=0\}}$ is an indicator function taking 1 when $y=0$ and 0 otherwise.

The set of premium levels is assumed to be $\textbf{c} =\{c_1, \ldots, c_5\}=\{11,12,14, 16, 18\}$ and the initial premium of new policyholders $C_1$ is $c_3$ that is 140\% of the expected aggregate reported claims $\mathbb{E}(X+Y)$ (i.e. a safety loading factor of 40\%). Under our assumption, the premium levels range from  $110\%$ to $180\%$ of the expected aggregate reported claims. We propose the following rules of premium adjustment:
\begin{enumerate}[label=\arabic*),nolistsep]
\item If the reported aggregate claims in the current period is no more than 3, then the premium level for the next period will move to the lower premium level or stay in the lowest one, i.e. for $s\le 3$, $t_{1 1}(s)=1, t_{i, i-1}(s)=1, i\ge 2$;
\item If the reported aggregate claims in the current period is more than 3 but no more than 14, then the premium level for the next period will remain in the current level, i.e. for $3< s\le 14$, $t_{i i}(s)=1, 1\le i\le 5$;
\item If the reported aggregate claims in the current period is more than 14, then the premium level for the next period will move to the higher premium level or stay in the highest one, i.e. for $s>14$, $t_{5 5}(s)=1, t_{i, i+1}(s)=1, i\le 4$.\\
\end{enumerate}

According to the above transition rules, we can calculate the transition probabilities among the premium levels based on \eqref{eq:2.5}. Let $\mathbf{P}^{H}_T$, $\mathbf{P}^{M}_T$ and $\mathbf{P}^{L}_T$ denote the transition matrix in each of the above correlation cases respectively, then we have
\begin{eqnarray*}
&&\mathbf{P}^{H}_T=\begin{bmatrix}0.76743&0.23257&0&0&0\\0.30556&0.46188&0.23257&0&0\\0&0.30556&0.46188&0.23257&0\\
0&0&0.30556&0.46188&0.23257\\0&0&0&0.30556&0.69444\end{bmatrix},\\
&&\mathbf{P}^{M}_T=\begin{bmatrix}0.75712&0.24288&0&0&0\\0.28407&0.47305&0.24288&0&0\\0 &0.28407&0.47305&0.24288&0\\
0&0&0.28407&0.47305&0.24288\\0&0&0&0.28407&0.71593\end{bmatrix},\\
&&\mathbf{P}^{L}_T=\begin{bmatrix}0.74681&0.25319&0&0&0\\0.26258&0.48423&0.25319&0& 0\\0 &0.26258&0.48423&0.25319&0\\
0&0&0.26258&0.48423&0.25319\\0&0&0&0.26258&0.73742\end{bmatrix}.
\end{eqnarray*}

\begin{figure}[H]
    \centering
    \includegraphics[scale=0.17]{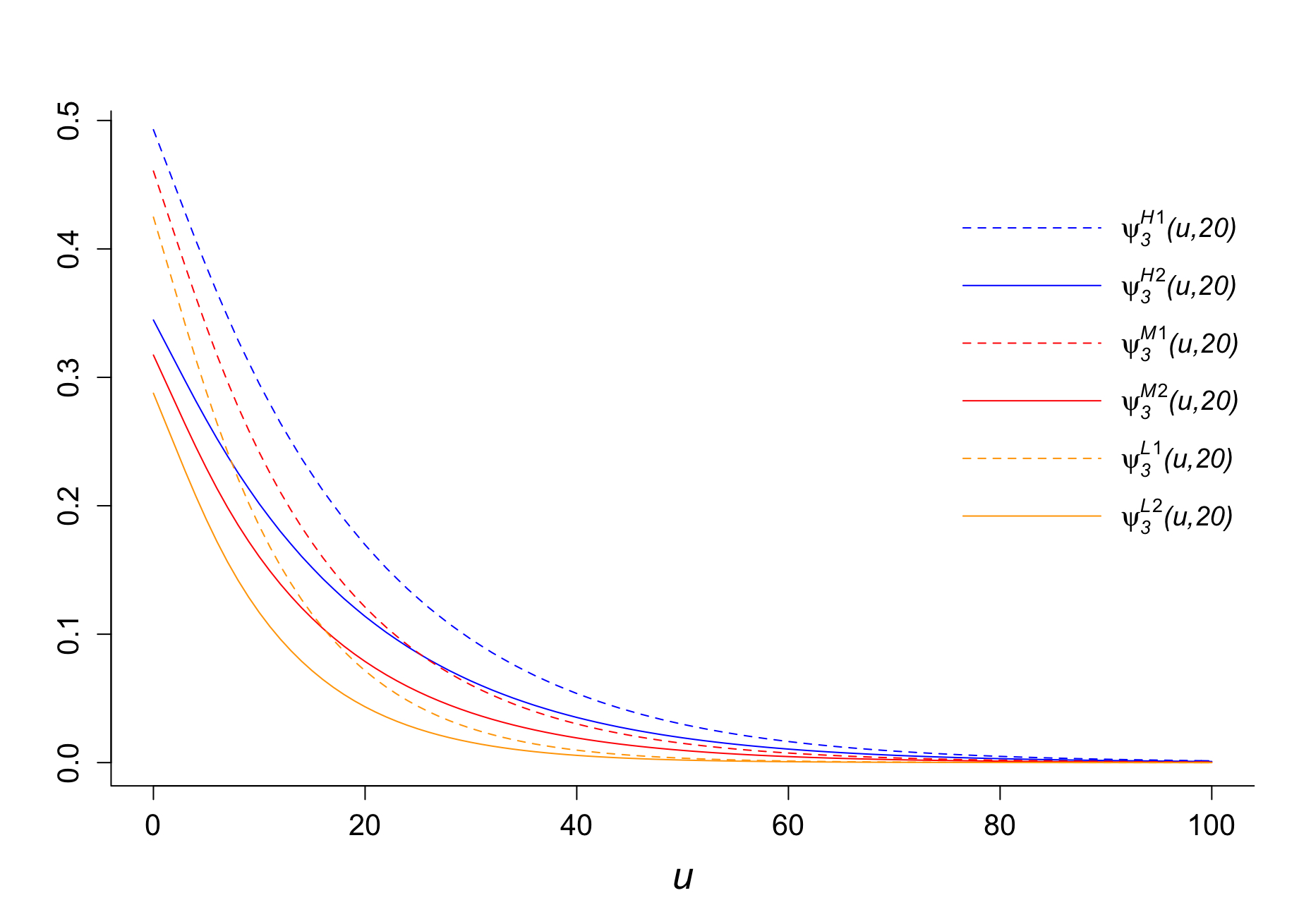}
    \caption{$\psi_{3}(u, 20)$ under the aggregate reported claims principle}
    \label{fig1}
\end{figure}

The corresponding long-term stationary distribution of the premium levels are: 
\begin{eqnarray*}
&&{\boldsymbol \pi^{H}}=[\pi^H_i]_{i\in\boldsymbol{\mathcal{L}}}=\begin{bmatrix} 0.32082&0.24419 &0.18586&0.14146&0.10767\end{bmatrix},\\
&&{\boldsymbol \pi^{M}}=[\pi^M_i]_{i\in\boldsymbol{\mathcal{L}}}=\begin{bmatrix} 0.26699&0.22828 &0.19518&0.16688&0.14268\end{bmatrix},\\
&&{\boldsymbol \pi^{L}}=[\pi^L_i]_{i\in\boldsymbol{\mathcal{L}}}=\begin{bmatrix}0.21482&0.20714 &0.19974&0.19259&0.18571\end{bmatrix}.
\end{eqnarray*}
The long-term expected premium per time period is 13.26, 13.65 and 14.07 in the high, moderate, and low correlation scenario, respectively. Using \eqref{eq:7} and \eqref{eq:8}, we calculate $\psi_3(u,20)$, $0\le u\le 100$, with the initial premium $c_3$ and the results are summarised in Table \ref{tab1} and Figure \ref{fig1}. Note that the notations $H1$, $M1$ and $L1$ denote the scenarios of high, moderate and low correlation between $X$ and $Y$ when $q=0.2$, and the notations $H2$, $M2$ and $L2$ correspond to the scenarios when $q=0.8$.
\medskip

\begin{longtable}{|c|c|c|c|c|c|c|}
\caption{$\psi_{3}(u, 20)$ under the aggregate reported claims principle}\label{tab1}\\
\hline
\multicolumn{1}{|c|}{\begin{tabular}[c]{@{}c@{}}$u$\end{tabular}} & \multicolumn{1}{c|}{\begin{tabular}[c]{@{}c@{}}$\psi^{H1}_{3}(u, 20)$\end{tabular}} & \multicolumn{1}{c|}{\begin{tabular}[c]{@{}c@{}}$\psi^{H2}_{3}(u, 20)$\end{tabular}} & \multicolumn{1}{c|}{\begin{tabular}[c]{@{}c@{}}$\psi^{M1}_{3}(u, 20)$\end{tabular}} & \multicolumn{1}{c|}{\begin{tabular}[c]{@{}c@{}}$\psi^{M2}_{3}(u, 20)$\end{tabular}} & \multicolumn{1}{c|}{\begin{tabular}[c]{@{}c@{}}$\psi^{L1}_{3}(u, 20)$\end{tabular}} & \multicolumn{1}{c|}{\begin{tabular}[c]{@{}c@{}}$\psi^{L2}_{3}(u, 20)$\end{tabular}}\\ \hline
\endhead
0 &0.48789&0.34433&0.46301&0.32119&0.43201&0.29416\\
10 &0.28527&0.19639&0.23543& 0.15643&0.17866&0.11266\\
20 &0.16386&0.11085&0.11795&0.07688&0.06897&0.04179\\  
30 &0.09279&0.06188&0.05892&0.03797&0.02564&0.01516\\ 
40 &0.05194&0.03423&0.02940&0.01878 &0.00931&0.00541\\ 
50 &0.02880&0.01878&0.01464&0.00929&0.00333&0.00191\\ 
60 &0.01583&0.01024&0.00728&0.00459&0.00117&0.00067\\ 
70 &0.00864&0.00554&0.00361&0.00226&0.00041&0.00023 \\ 
80 &0.00469&0.00298&0.00178&0.00111&0.00014&0.00008 \\ 
90 &0.00253&0.00160&0.00088&0.00054 &0.00005&0.00003 \\ 
100 &0.00136&0.00085&0.00043&0.00027&0.00002&0.00001\\\hline
\end{longtable}

The first observation, a trivial one, from Table \ref{tab1} and Figure \ref{fig1} is that $\psi_3(u,20)$ decreases when $u$ increases. Moreover, we notice that the correlation level between main claim $X$ and by-claim $Y$ does affect the finite-time ruin probability. Under our previous assumptions, after fixing $u$ and $q$, the higher is the correlation, the higher is the risk of ruin. Although the same premium adjustment rules are applicable for all three correlation scenarios, the joint distribution of $X$ and $Y$ differentiates the transition probabilities among premium levels as well as the stationary distribution of individual premium levels. The previously calculated $\pi^H$, $\pi^M$ and $\pi^L$ show that the high correlation case has the highest long-term probability to reach low premium levels and the lowest long-term probability for high premium levels. It implies that in long-run, in scenario $H$, the insurer is expected to receive less total premium income than the other two scenarios, which results in the highest finite-time ruin probabilities among the three scenarios. Similar arguments can be made to explain the ordering between cases $M$ and $L$.  

In addition, the differences, in terms of percentages, among the finite-time ruin probabilities under the three scenarios increase when the initial surplus $u$ increases. For example, $\psi^{H1}_{3}(0, 20)$ is only about 5.4\% higher than $\psi^{M1}_{3}(0, 20)$ and around 12.9\% higher than $\psi^{L1}_{3}(0, 20)$. But $\psi^{H1}_{3}(100, 20)$ is about three times $\psi^{M1}_{3}(100, 20)$ and around  68 times $\psi^{L1}_{3}(100, 20)$. This makes sense because when $u$ is small, if ruin occurs then it is more likely to occur within the first few periods. As a result, there is only limited time for the main factors, which vary the finite-time ruin probabilities among these scenarios, to take effect. The same initial premium assumption under all three scenarios also contributed to the small differences in percentage among the finite-time ruin probabilities when $u$ is small. On the contrary, when $u$ is large, if ruin occurs then ruin is more likely to occur in the long run. The dissimilar premium evolving patterns under the three scenarios have plenty of time to drive the underlying surplus processes to different directions, which lead to divergent finite-time ruin probabilities. 

Last but not least, it is evident from Figure \ref{fig1} that with all other factors being the same, an increase in $q$ from 0.2 to 0.8 shifted the finite-time ruin probabilities downwards. This is reasonable since when the settlement of by-claims is more likely to be delayed, the insurers can receive more premium income that helps to settle the claims. However, this effect reduces when $u$ is larger, because delaying by-claims for one time unit would not make a big difference for the worst cases (i.e. getting bankrupted with a large initial capital).

\subsection{Premiums adjusted according to aggregate settled claims}

This example examines the premium adjustment principle that was discussed in Section 4. This principle is worth exploring because that, in certain circumstances, the total settled claim amounts might better reflect the claims experiences of policyholders than the total reported claim amounts in a given time window due to the fact that in real practice reported claims come with uncertainties in the scale and timing of the real settlements. Therefore, the reported claims are only initial guesses and may not provide accurate information to represent the policyholders' historical claim experience. In this paper, for the purpose of simplification, we assumed that the reported and settled by-claims amounts are always equal and the length of delay is always 1. Although these restrictive assumptions are not entirely realistic, they serve as good starting points that could motivate more realistic models in future studies. \medskip

We assume the same claim distributions and premium levels as those in previous example, whilst the transition rules of premium levels are modified as:\\
\begin{enumerate}[label=\arabic*),nolistsep]
\item If the settled aggregate claims in the current period is no more than 3, the premium level for the next period will move to the lower premium level or stay in the lowest one;
\item If the settled aggregate claims in the current period is more than 3 but no more than 14, the premium level for the next period will remain in the current premium level;
\item If the settled aggregate claims in the current period is more than 14, the premium level for the next period will move to the higher premium level or stay in the highest one.\\
\end{enumerate}

\begin{figure}[H]
    \centering
    \includegraphics[scale=0.17]{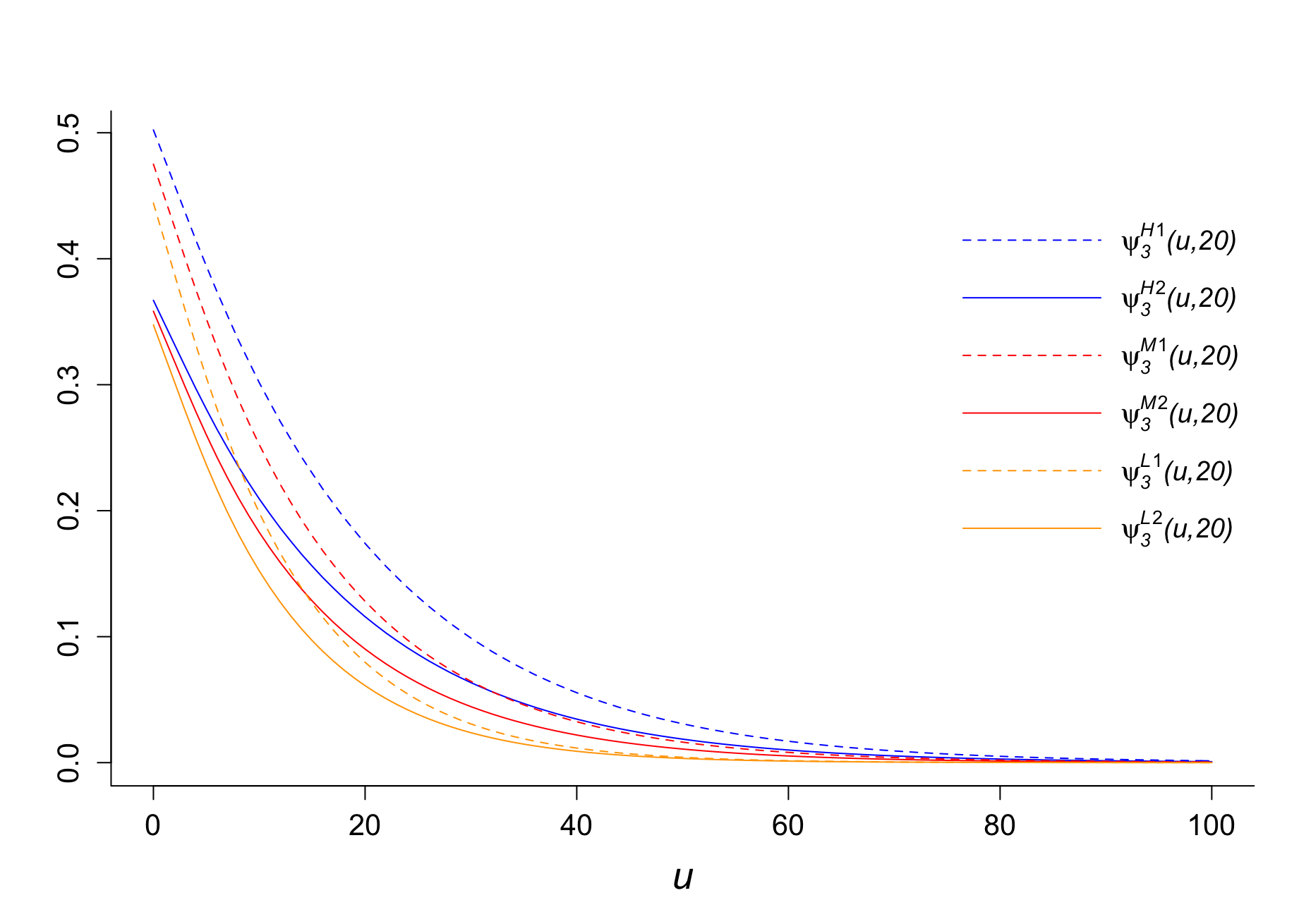}
    \caption{$\psi_{3}(u, 20)$ under the aggregate settled claims principle}
    \label{fig2}
\end{figure}

By the non-homogeneity nature exhibited under the new rules, there is no constant one-step transition matrix among the premium levels anymore. On the contrary, the one-step transition matrix varies over time and depends on the number of by-claims settled in each given time period. However, we can still study 20-period finite-time ruin probabilities using the recursive formulae \eqref{eq:9} and \eqref{eq:10}. The results are given in Table \ref{tab2} and Figure \ref{fig2}. We adopt the same notation to denote the scenarios under consideration. 

\begin{longtable}{|c|c|c|c|c|c|c|}
\caption{$\psi_{3}(u, 20)$ under the aggregate settled claims principle}\label{tab2}\\
\hline
\multicolumn{1}{|c|}{\begin{tabular}[c]{@{}c@{}}$u$\end{tabular}} & \multicolumn{1}{c|}{\begin{tabular}[c]{@{}c@{}}$\psi^{H1}_{3}(u, 20)$\end{tabular}} & \multicolumn{1}{c|}{\begin{tabular}[c]{@{}c@{}}$\psi^{H2}_{3}(u, 20)$\end{tabular}} & \multicolumn{1}{c|}{\begin{tabular}[c]{@{}c@{}}$\psi^{M1}_{3}(u, 20)$\end{tabular}} & \multicolumn{1}{c|}{\begin{tabular}[c]{@{}c@{}}$\psi^{M2}_{3}(u, 20)$\end{tabular}} & \multicolumn{1}{c|}{\begin{tabular}[c]{@{}c@{}}$\psi^{L1}_{3}(u, 20)$\end{tabular}} & \multicolumn{1}{c|}{\begin{tabular}[c]{@{}c@{}}$\psi^{L2}_{3}(u, 20)$\end{tabular}}\\ \hline
\endhead
0 &0.49739&0.36760&0.47738&0.36262&0.45114&0.35399\\
10 &0.29196& 0.20393&0.24635&0.17862&0.19275 &0.14766\\  
20 &0.16826& 0.11276 & 0.12495&0.08811&0.07701&0.05910 \\ 
30 &0.09555&0.06178 &0.06303 &0.04346&0.02963&0.02294 \\ 
40&0.05361&0.03358&0.03170&0.02143&0.01112&0.00869 \\ 
50 &0.02978&0.01813 & 0.01590&0.01056 &0.00410&0.00323  \\ 
60 &0.01640&  0.00974 &0.00795&0.00519&0.00149&0.00118   \\ 
70 &0.00896&0.00520&0.00396& 0.00254&0.00053&0.00043  \\ 
80 &0.00487& 0.00277&0.00196& 0.00125&0.00019&0.00015  \\ 
90 &0.00263& 0.00147&0.00097&0.00061&0.00007&0.00005  \\ 
100&0.00141&0.00077&0.00048&0.00030&0.00002&0.00002\\\hline
\end{longtable}

As shown in Table \ref{tab2} and Figure \ref{fig2}, consistent observations are evident in this aggregate settled claims principle comparing with the aggregate reported claims case. Further, the differences between the two $q$ cases in each correlation scenario also behave interestingly differently. In the high correlation scenario, there is a big gap between the two ruin probability curves showing that a high chance of delaying the highly correlated by-claims results in a big reduction in the risk of ruin comparing from the case of low chance of delay. On the contrary, when the correlation between main claims and by-claims is low and $u$ is not small, whether delaying the by-claims or not seem not having a significant impact on the finite-time ruin probabilities. A reasonable interpretation is that when the correlation is low, the main difference between the two cases of $q$ is that the by-claims settled in each time period are likely to be delayed ones or freshly incurred ones. Since the correlation between the main claims and by-claims is low, the distributions of aggregate settled claims in each period are similar in both cases. Therefore, except the first time period, the surplus process should behave similarly within all remaining time periods in both $q$ cases that lead to similar finite-time ruin probabilities.\medskip

\begin{figure}[H]
    \centering
    \includegraphics[scale=0.17]{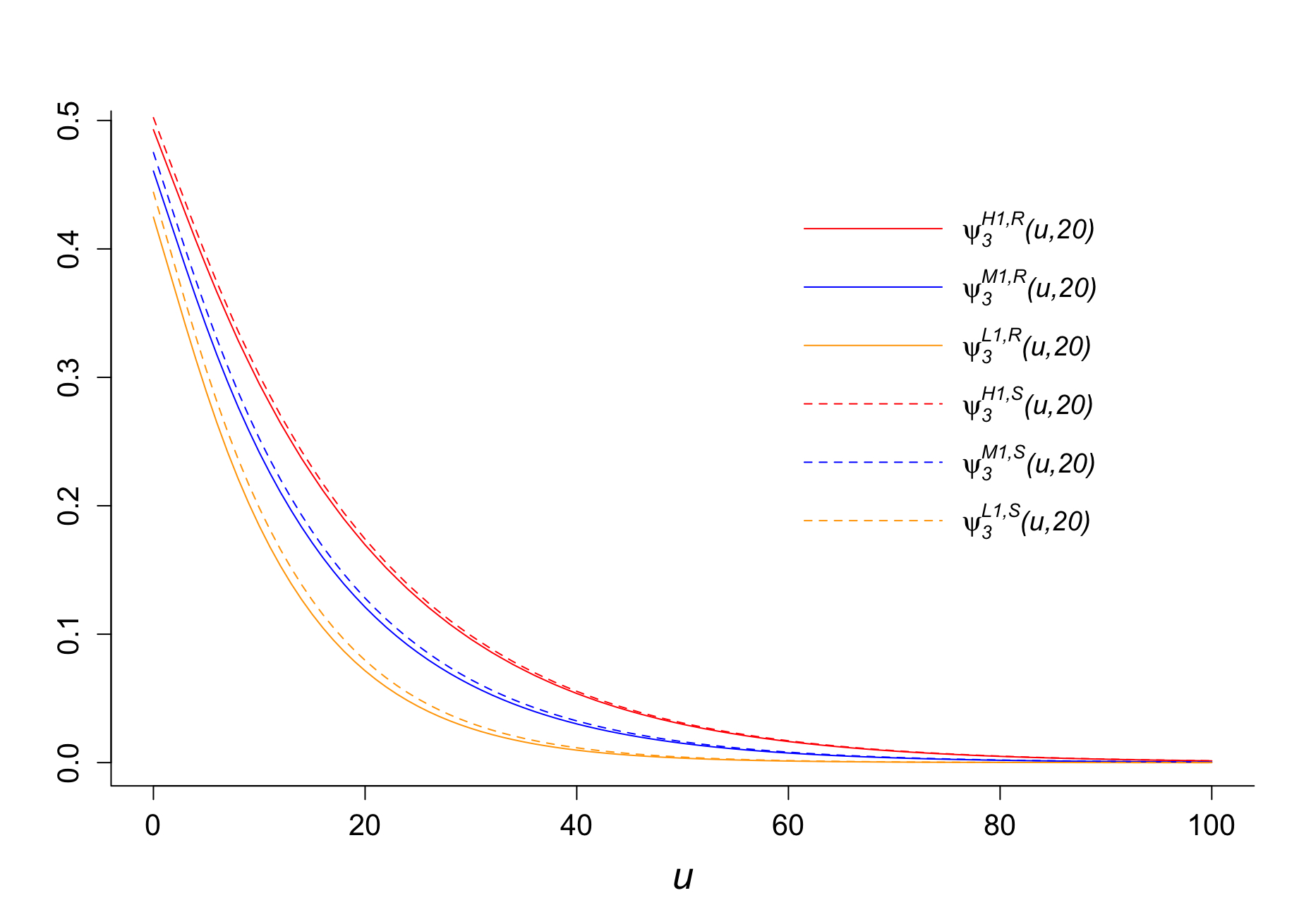}
    \caption{Comparison between $\psi_{3}(u, 20)$ in 7.1 and 7.2 when $q=0.2$.}
    \label{fig2a}
\end{figure}

\begin{figure}[H]
    \centering
    \includegraphics[scale=0.17]{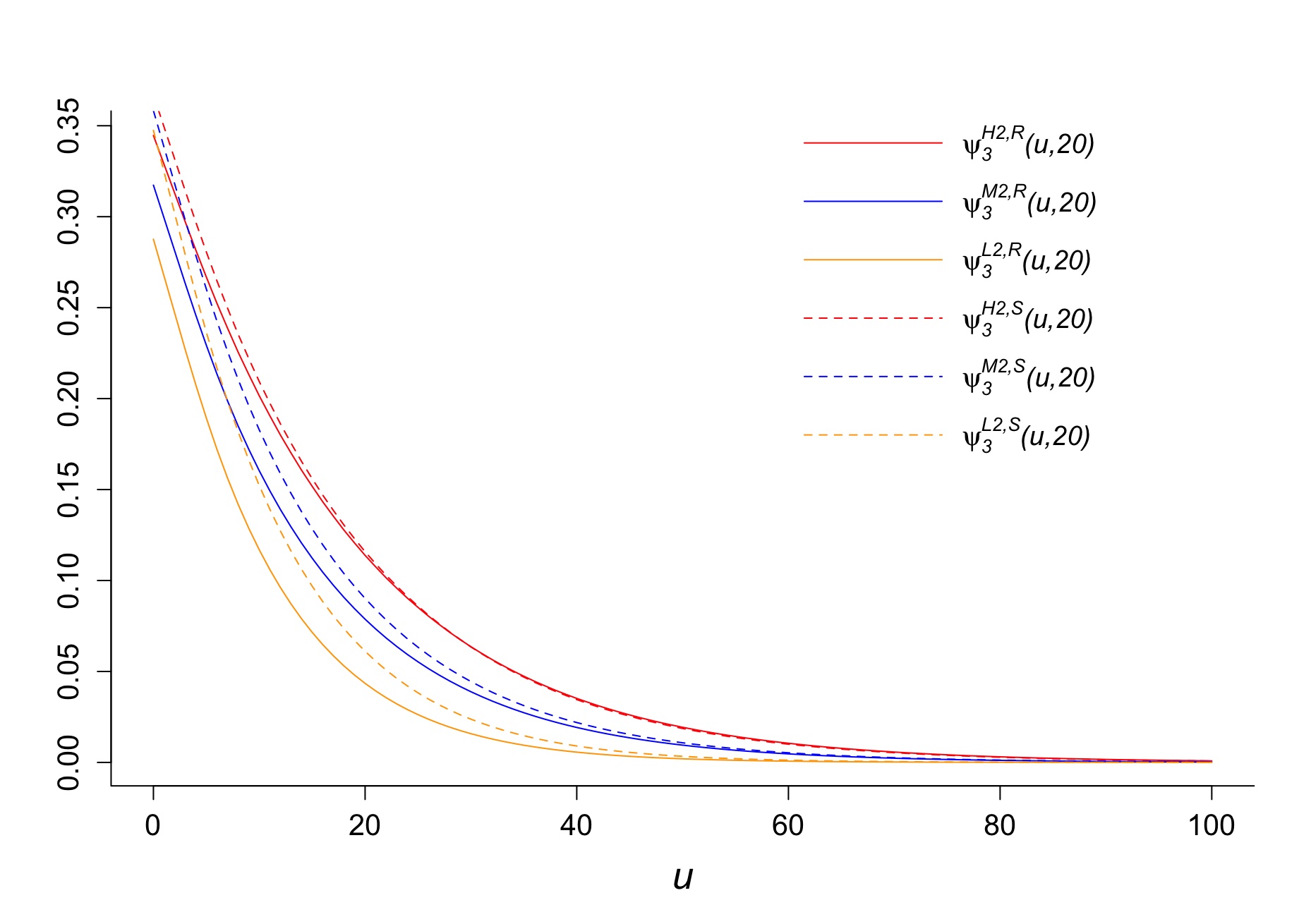}
    \caption{Comparison between $\psi_{3}(u, 20)$ in 7.1 and 7.2 when $q=0.8$.}
    \label{fig2b}
\end{figure}

Moreover, we generate comparison results, shown in Figure \ref{fig2a} and Figure \ref{fig2b}, regarding $\psi_3(u, 20)$ in this and the previous numerical examples. The superscripts $R$ and $S$ denote the premium adjustment principle by reported aggregate claims and by settled aggregate claims respectively. 

As seen in Figure \ref{fig2a}, the two premium correction principles lead to marginal differences in the finite-time ruin probabilities in the case of $q=0.2$, because when $q$ is small, the aggregate reported claims in each period are likely to be the same as the aggregate settled claims. Therefore, the periodic premiums are highly likely to follow the same pattern in both cases, which result in similar finite-time ruin probabilities. \medskip

On the other hand, according to Figure \ref{fig2b}, when $q=0.8$ the trends of finite-time ruin probabilities in the two cases differ significantly from one another. However, the differences increase when the correlation between the main claims and by-claims becomes weaker, and they tend to diminish when $u$ increases. Moreover, when $q=0.8$ the differences among the three correlation scenarios under the aggregate settled claims principle are generally smaller than those in the aggregate reported claims case. A possible interpretation is that when $q$ is high, after the first couple of time periods, the aggregate settled claims in each period is highly likely to be the summation of a main claim $X$ of the current period and a by-claim $Y$ delayed from the previous period (if any), whilst the aggregate reported claims in each period is a current main claim plus a current by-claim (if any). Due to the independence assumption between main claims and by-claims in different time periods, the within-period correlation between main and by-claims becomes between-period correlation in the aggregate settled claims case, which likely contributes to the above observation.    \medskip

A consistent finding in both $q$ cases is that the finite-time ruin probabilities under the aggregated settled claims principle are generally higher than the corresponding ones in the aggregate reported claims case. It implies that if the information regarding reported claims is accurate, then the insurers better adopt the aggregate reported claims principle to adjust their periodic premiums, or they will face a higher insolvency risk otherwise. \medskip

In the following sections, we shall provide two examples designed to examine the finite-time ruin probabilities with premiums adjustment principles that focus on the claim frequency information.

\subsection{Premiums adjusted according to reported claims number}

In this example, we assume that the claim distributions and the set of premium levels are the same as the previous examples. The rules of premium corrections are:\medskip

\begin{enumerate}[label=\arabic*),nolistsep]
\item If the number of reported claims in the current period is 0, then the premium level for the next period will move to the lower premium level or stay in the lowest one;
\item If the number of reported claims in the current period is 1, then the premium level for the next period will remain in the current premium level;
\item If the number of reported claims in the current period is more than 1, then the premium level for the next period will move to the higher premium level or stay in the highest one.\\
\end{enumerate}

Next, we shall explore the impact of the correlation between main claims and by-claims as well as the impact of $q$ on the finite-time ruin probabilities. Under the new premium adjustment rules given above, the correlation between the number of main claims $N^X_t$ and by-claims $N^Y_t$ are calculated instead of the correlation between $X$ and $Y$. We find that the correlation between $N^X_t$ and $N^Y_t$ generated by the claims distribution $f^{H}_{X Y}(x,y)$, $f^{M}_{X Y}(x,y)$ and $f^{L}_{X Y}(x,y)$ is $\rho_{N^X, N^Y}=1$,  $\rho_{N^X, N^Y}=0.8272$ and $\rho_{N^X, N^Y}=0.7071$, respectively. These surprisingly high correlations between number of claims are rooted in the model assumptions made in Section 2, i.e. one main claim generates at most one by-claim and no main claim means no by-claim. \medskip

\begin{figure}[H]
    \centering
    \includegraphics[scale=0.17]{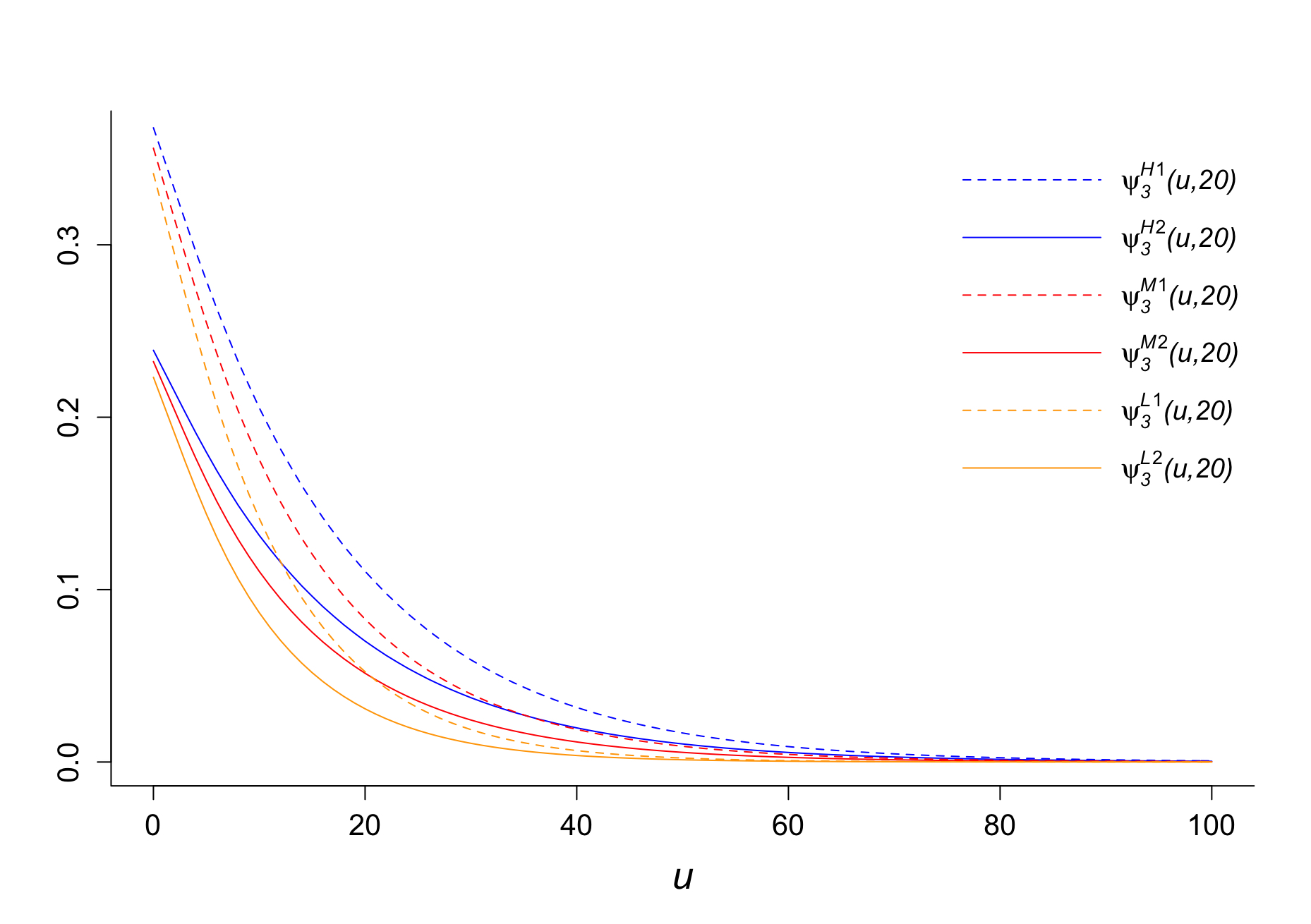}
    \caption{$\psi_{3}(u, 20)$ under the reported claims number principle}
    \label{fig3}
\end{figure}

Similar to the example in Section 7.1, we can calculate the transition matrix among premium levels as follows:

\begin{eqnarray*}
\mathbf{P}^{H}_T= \begin{bmatrix}1/6&5/6&0&0&0\\1/6&0&5/6&0&0\\0&1/6 &0&5/6&0\\
0&0&1/6&0&5/6\\0&0&0&1/6&5/6\end{bmatrix},
\end{eqnarray*}
\begin{eqnarray*}
\mathbf{P}^{M}_T= \begin{bmatrix}0.22619&0.77381&0&0&0\\0.16667&0.05952&0.77381&0&0\\0& 0.16667&0.05952&0.77381&0\\
0&0&0.16667&0.05952&0.77381\\0&0&0&0.16667&0.83333\end{bmatrix},
\end{eqnarray*}
\begin{eqnarray*}
\mathbf{P}^{L}_T= \begin{bmatrix}0.28571&0.71429&0&0&0\\0.16667&0.11905&0.71429&0&0\\0&0.16667& 0.11905&0.71429&0\\
0&0&0.16667&0.11905&0.71429\\0&0&0&0.16667&0.83333\end{bmatrix}.
\end{eqnarray*}
The corresponding long-term stationary distribution of the premium levels are:
\begin{eqnarray*}
{\boldsymbol \pi^{H}}&=& \begin{bmatrix}0.00128&0.00640&0.03201&0.16005 &0.80026\end{bmatrix},\\
{\boldsymbol \pi^{M}}&=& \begin{bmatrix}0.00169&0.00784&0.03642&0.16907 &0.78498\end{bmatrix},\\
{\boldsymbol \pi^{L}}&=& \begin{bmatrix}0.00227&0.00975&0.04177&0.17901 &0.76720\end{bmatrix}.
\end{eqnarray*}
The long-term expected premiums in each correlation scenario is 17.50, 17.46 and 17.40 in the H, M and L scenario, respectively. It is worth noting that given the very different joint distributions of $X$ and $Y$ in the three correlation scenarios, the corresponding long-term expected premiums are very similar under the current premium correction principle. By \eqref{eq:11} and \eqref{eq:12}, we obtain results for $\psi_3(u,20)$ that are summarized in Table \ref{tab3} and Figure \ref{fig3}. The notations in Table \ref{tab3} are defined in the same way as in Table \ref{tab1} and \ref{tab2}.

\begin{longtable}{|c|c|c|c|c|c|c|}
\caption{$\psi_{3}(u, 20)$ under the reported claims number principle\label{tab3}}\\
\hline
\multicolumn{1}{|c|}{\begin{tabular}[c]{@{}c@{}}$u$\end{tabular}} & \multicolumn{1}{c|}{\begin{tabular}[c]{@{}c@{}}$\psi^{H1}_{3}(u, 20)$\end{tabular}} & \multicolumn{1}{c|}{\begin{tabular}[c]{@{}c@{}}$\psi^{H2}_{3}(u, 20)$\end{tabular}} & \multicolumn{1}{c|}{\begin{tabular}[c]{@{}c@{}}$\psi^{M1}_{3}(u, 20)$\end{tabular}} & \multicolumn{1}{c|}{\begin{tabular}[c]{@{}c@{}}$\psi^{M2}_{3}(u, 20)$\end{tabular}} & \multicolumn{1}{c|}{\begin{tabular}[c]{@{}c@{}}$\psi^{L1}_{3}(u, 20)$\end{tabular}} & \multicolumn{1}{c|}{\begin{tabular}[c]{@{}c@{}}$\psi^{L2}_{3}(u, 20)$\end{tabular}}\\ \hline
\endhead
0 &0.36310&0.23848&0.35810&0.23559 &0.34799 & 0.22890\\
10&0.19645 &0.12700& 0.16968 &0.10723&0.13642 & 0.08316 \\  
20 &0.10571&0.06772 &0.08018&0.05000 &0.05032&0.02958\\ 
30 & 0.05661&0.03601& 0.03820&0.02369 &0.01801& 0.01038\\ 
40 &0.03020& 0.01910& 0.01834 &0.01134&0.00634& 0.00361\\ 
50 & 0.01606 & 0.01011&0.00885&0.00546&0.00221& 0.00125\\ 
60&0.00852&0.00535& 0.00428&0.00263&0.00076 & 0.00043  \\ 
70& 0.00451& 0.00282 & 0.00208 & 0.00127 &0.00026 &0.00015 \\ 
80 & 0.00238&0.00149 &0.00101&0.00062&0.00009& 0.00005  \\ 
90& 0.00126&0.00078 & 0.00049& 0.00030 &0.00003&0.00002  \\ 
100& 0.00066&0.00041& 0.00024&0.00014&0.00001&0.00001\\\hline
\end{longtable}

Again, Table \ref{tab3} and Figure \ref{fig3} show us some similar trends to those shown in Table \ref{tab1} \& \ref{tab2} and Figure \ref{fig1} \& \ref{fig2}. First, $\psi_3(u,20)$ decreases when $u$ increases and the correlation level between $N^X$ and $N^Y$ is positively related to the ruin probabilities. When fixing $u$ and $q$, the higher the correlation, the higher is the ruin probability. Additionally, the decrease in $q$ from 0.8 to 0.2 also causes a lift in the finite time ruin probabilities in all correlation scenarios. There are two inconsistencies between this example and the previous ones: 
\begin{itemize}
\item Firstly, the scales of difference in $\rho_{X, Y}$ and $\psi_3(u, 20)$ among all correlation scenarios in Section 7.1 and 7.2 are larger than the corresponding differences in this example. An interpretation is that, as given at the beginning of this section, the differences among the three $\rho_{N^X, N^Y}$ values are much smaller than the differences among the three $\rho_{X, Y}$ values, which makes the three correlation scenarios less distinct from one another.
\item Secondly, the relationship between $\rho_{N^X, N^Y}$ and the long-term expected premium in this example is opposite to that in Section 7.1. To be more specific, in Section 7.1, lower $\rho_{X, Y}$ leads to higher long-term expected premiums, whereas in this example, lower $\rho_{N^X, N^Y}$ gives lower long-term expected premiums. A likely justification of this difference is the change of premium correction objectives from aggregate claim experience to claim frequencies. 
\end{itemize}

\subsection{Premiums adjusted according to settled claims number}

In our last numerical example, we shall duplicate the model assumptions but change the premiums adjustment rules following the settled claims number premium principle. The transition rules of premiums are:
\begin{enumerate}[label=\arabic*)]
\item If the number of settled claims in the current period is 0, then the premium level for the next period will move to the lower premium level or stay in the lowest one;
\item If the number of settled claims in the current period is 1, then the premium level for the next period will remain in the current premium level;
\item If the number of settled claims in the current period is more than 1, the premium level for the next period will move to the higher premium level or stay in the highest one.\\
\end{enumerate}

\begin{figure}[H]
    \centering
    \includegraphics[scale=0.17]{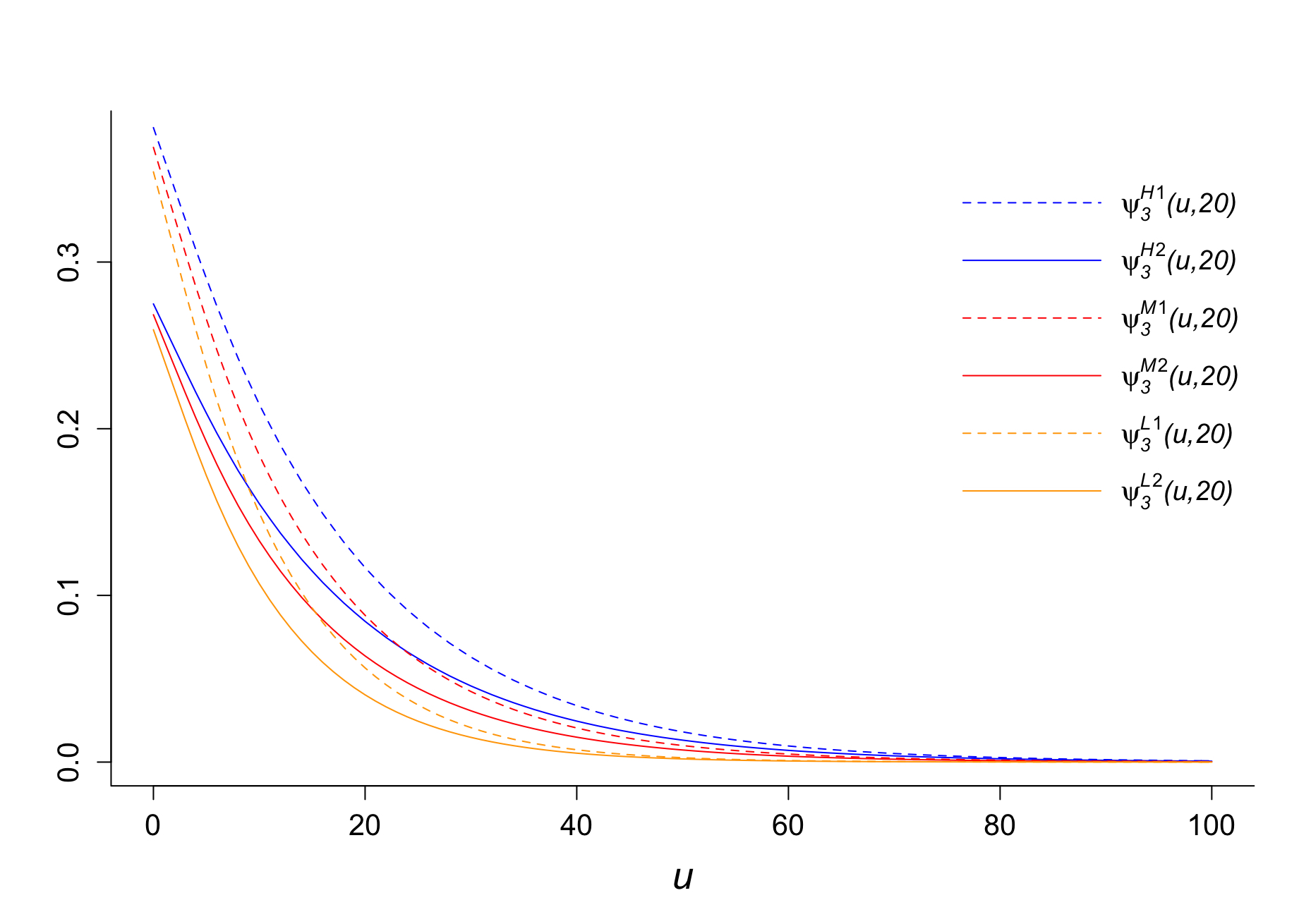}
    \caption{$\psi_{3}(u, 20)$ under the settled claims number principle}
    \label{fig4}
\end{figure}

We use \eqref{eq:13} and \eqref{eq:14} to calculate $\psi_3(u,20)$ and the results are summarised in Table \ref{tab4} and Figure \ref{fig4}, adopting the same notations. 

\begin{longtable}{|l|l|l|l|l|l|l|}
\caption{$\psi_{3}(u, 20)$ under the settled claims number principle\label{tab4}}\\
\hline
\multicolumn{1}{|c|}{\begin{tabular}[c]{@{}c@{}}$u$\end{tabular}} & \multicolumn{1}{c|}{\begin{tabular}[c]{@{}c@{}}$\psi^{H1}_{3}(u, 20)$\end{tabular}} & \multicolumn{1}{c|}{\begin{tabular}[c]{@{}c@{}}$\psi^{H2}_{3}(u, 20)$\end{tabular}} & \multicolumn{1}{c|}{\begin{tabular}[c]{@{}c@{}}$\psi^{M1}_{3}(u, 20)$\end{tabular}} & \multicolumn{1}{c|}{\begin{tabular}[c]{@{}c@{}}$\psi^{M2}_{3}(u, 20)$\end{tabular}} & \multicolumn{1}{c|}{\begin{tabular}[c]{@{}c@{}}$\psi^{L1}_{3}(u, 20)$\end{tabular}} & \multicolumn{1}{c|}{\begin{tabular}[c]{@{}c@{}}$\psi^{L2}_{3}(u, 20)$\end{tabular}}\\ \hline
\endhead
0&0.37559 &0.27392 & 0.37074&0.27144 &0.36068 &0.26506\\
10&0.20550&0.15024 & 0.17838&0.12923 &0.14449& 0.10328\\
20& 0.11160&0.08175& 0.08534&0.06204& 0.05439& 0.03884\\  
30& 0.06024& 0.04420 & 0.04106 & 0.02999& 0.01984& 0.01424 \\ 
40&0.03236&0.02376&0.01986& 0.01456 & 0.00710&0.00513\\ 
50&0.01731& 0.01272& 0.00964&0.00709& 0.00251& 0.00182\\ 
60&0.00923& 0.00678& 0.00469 & 0.00345 & 0.00088 & 0.00064 \\ 
70& 0.00491& 0.00360&0.00228 &0.00168 &0.00030&0.00022\\ 
80&0.00260& 0.00191 &0.00111&0.00082 & 0.00011& 0.00008 \\ 
90& 0.00138&0.00101& 0.00054&0.00040&0.00004 & 0.00003 \\ 
100&0.00073& 0.00053& 0.00026& 0.00019&0.00001& 0.00001\\\hline
\end{longtable}

\begin{figure}[H]
    \centering
    \includegraphics[scale=0.17]{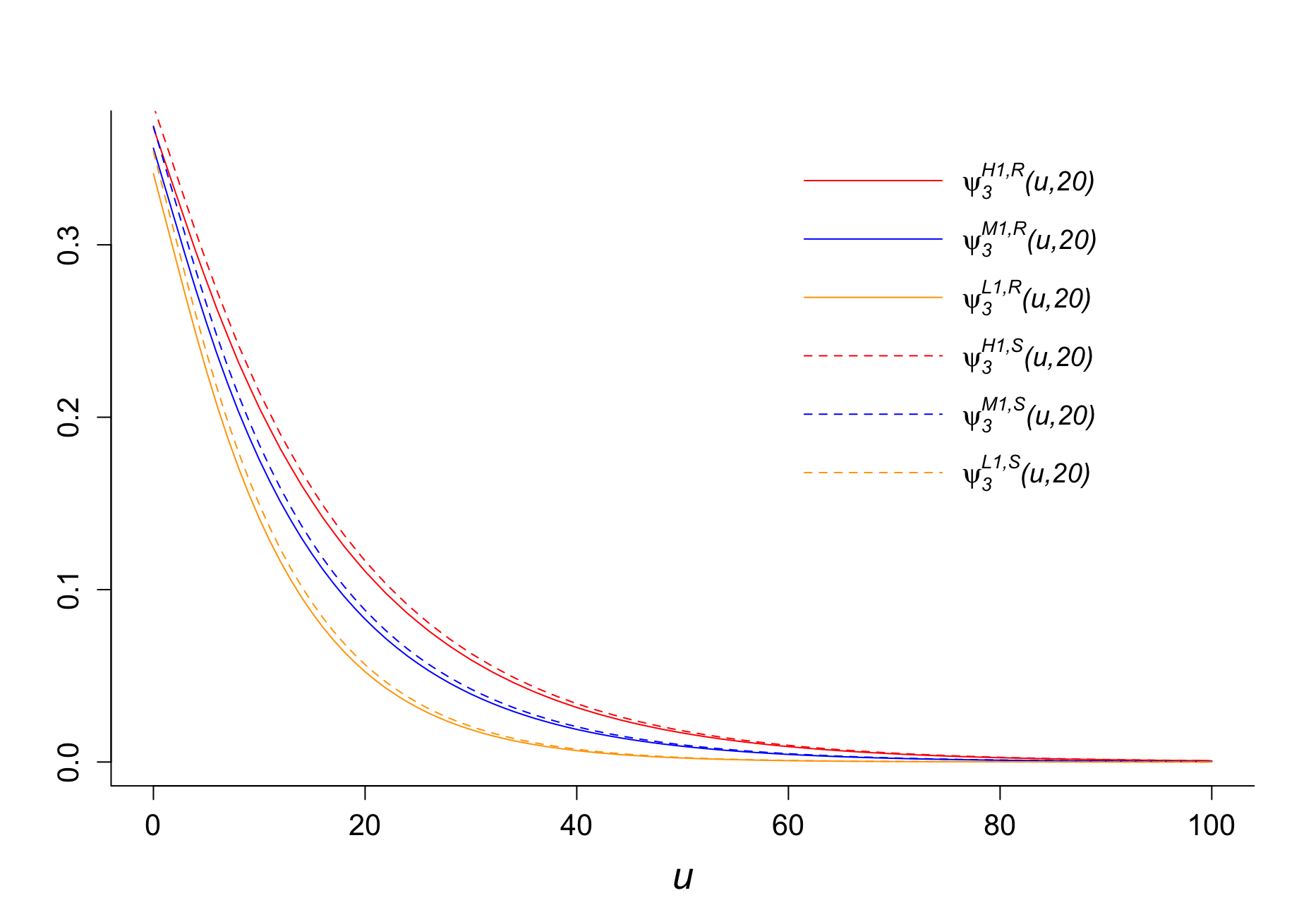}
    \caption{Comparison between $\psi_{3}(u, 20)$ in 7.3 and 7.4 when $q=0.2$.}
    \label{fig4a}
\end{figure}

\begin{figure}[H]
    \centering
    \includegraphics[scale=0.17]{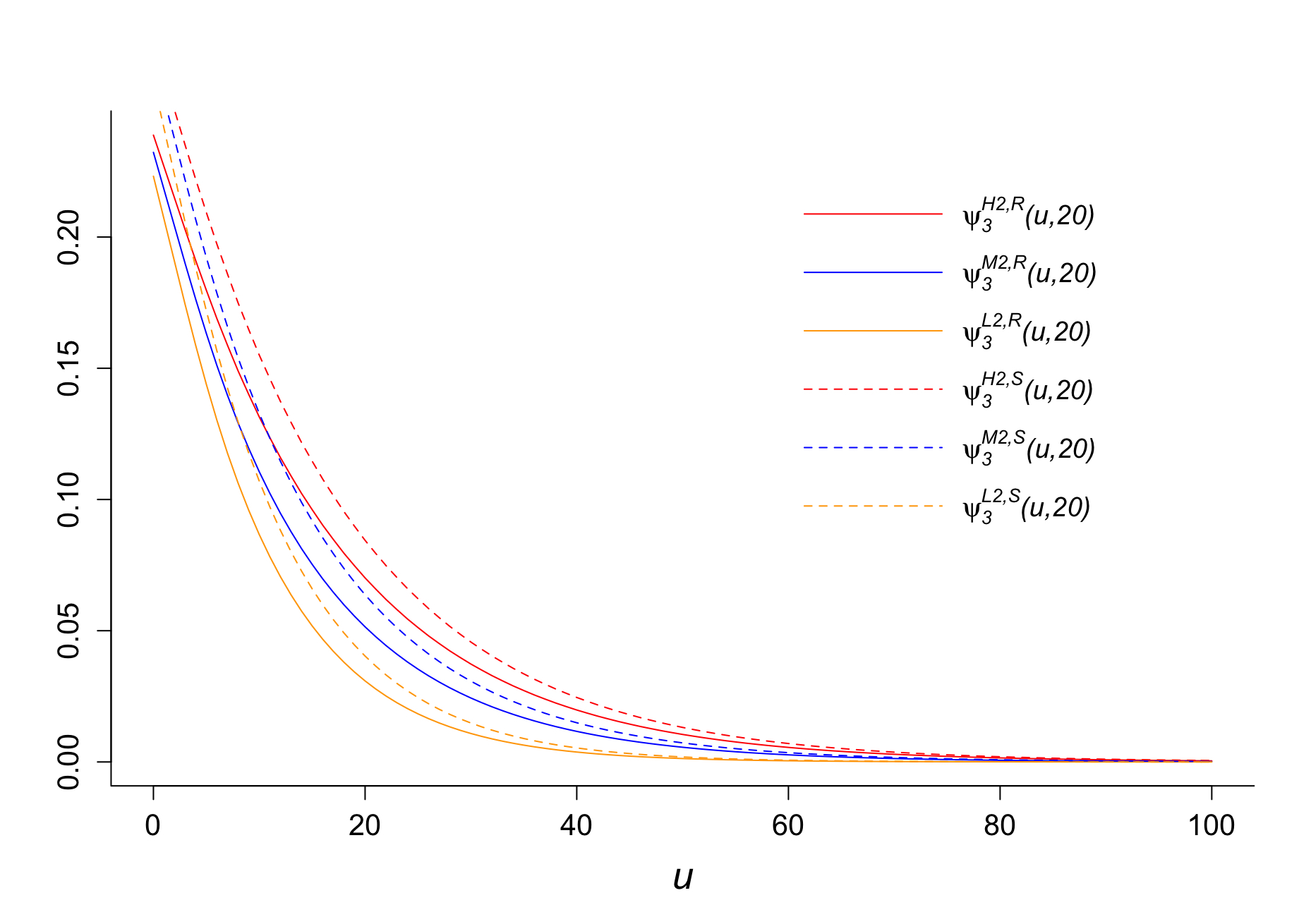}
    \caption{Comparison between $\psi_{3}(u, 20)$ in 7.3 and 7.4 when $q=0.8$.}
    \label{fig4b}
\end{figure}

Table \ref{tab4} and Figure \ref{fig4} show very similar trends to our findings from Table \ref{tab3} and Figure \ref{fig3}. Again, we generate two comparison graphs, Figure \ref{fig4a} and Figure \ref{fig4b}, between the two claim frequency premium principles for $q=0.2$ and $q=0.8$ respectively. The superscript $R$ denotes the reported claims number principle and $S$ denotes the settled claims number one. 

From Figure \ref{fig4a} and Figure \ref{fig4b}, we can see that when $q=0.2$, the finite-time ruin probabilities in this example is slightly higher than the results in section 7.3 for fixed $u$ and the correlation level. On the other hand, when $q=0.8$, the gaps between the finite-time ruin probabilities of this example and those of 7.3 are larger. This is because when $q=0.2$, the by-claim settlements are unlikely to be delayed. Therefore, it is more likely that both main claims and their associated by-claims to be settled in the same time periods, which makes the reported claims number principle and the settled claims number one to work similarly. On the contrary, when $q=0.8$, the settled claims number in the first time period is likely to be one since there is no up-front delayed by-claim, while the by-claim's settlement (if any) is likely to be delayed. Therefore, the settled claims number principle would determine the second premium according to the number of main claims in period one, whereas both the number of main claims and by-claims in period one will be used by the reported claims number principle. As a result, it is likely that the second premium under the reported claims number principle will be higher than the one under the settled claims number principle, which varies the whole sequence of future premiums and results in lower ruin probabilities in the former case.

\section{Concluding remarks and future research}
In this paper, we studied a discrete-time risk model with claim-dependent premiums and time-delayed by-claims. Our main goal is to evaluate the impact of the correlation between the main claims and by-claims and the probability of delaying by-claim settlements on the finite-time ruin probabilities under the proposed premium adjustment principles: the aggregate reported claims principle, the aggregate settled claims principle, the reported claims number principle and the settled claims number principle. Under certain assumptions, we found in our numerical studies that a higher probability of delaying the by-claim settlements would result in lower finite-time ruin probabilities. Moreover, the higher correlation between the main claims and by-claims also leads to higher finite-time ruin probabilities. Lastly, the premium adjustment principles based on settled claims experience (aggregate settled claims or settled claims number) account for higher finite-time ruin probabilities, compared with the principles based on the reported claims experience, given all other factors are the same. This difference is more remarkable when the probability of by-claim delays is high. According to these main findings in our study, the insurers should remain on high alert if a high correlation between the main claims and their associated by-claims is evident or the chance of getting delays in claim settlements is low because both situations could lead to increased insolvency risk. Further, the premium adjustment principles based on the reported claims experience could be a safer choice than the principles based on settled claims experience, especially in the high probability of delayed by-claim settlement cases. \medskip

However, there are some limitations in this study that could be addressed in future research. Firstly, the numerical results of this study only assumed a positive correlation between the main claims and by-claims, but in real life, the correlation can also be negative. Secondly, this paper assumed that by-claim settlements could only be delayed by one time period, which is not realistic in real practice. As an extension, a multiple-period delay could be taken into consideration. Finally, our study assumed that there were at most one main claim and one by-claim incurring in each period and the settled claim amounts are always equal to the reported ones. Some more realistic models allowing general main claim and/or by-claim counts, as well as unequal amounts in reporting and settlement might be worth studying in future research.




\end{document}